\documentclass{cernrep}

\usepackage{siunitx}
\usepackage{cite}
\usepackage[bookmarks, colorlinks=true, linktoc=page, linkcolor=black, citecolor=black, urlcolor=blue]{hyperref}
\usepackage{subfigure}
\usepackage{float}
\usepackage{fancyhdr}
\fancyhfoffset{4 mm}
\fancypagestyle{ARTTITLE}{%
\fancyhf{} 
\lhead{\footnotesize{Proceedings of the 2019 CERN--Accelerator--School course on
\it{ High Gradient Wakefield Accelerators}, Sesimbra, (Portugal)}}
\lfoot{Available online at \url{https://cas.web.cern.ch/previous-schools}}
\rfoot{\thepage\hspace*{3mm}}

   }
\begin{document}
\title{Plasma Sources and Diagnostics}
\author{M.~J.~Garland, J.~C.~Wood, G.~Boyle, and J.~Osterhoff}
\institute{Deutsches Elektronen-Synchrotron DESY, Hamburg, Germany}


\begin{abstract}
Carefully engineered, controlled, and diagnosed plasma sources are a key ingredient in mastering plasma-based particle accelerator technology. This work reviews basic physics concepts, common types of plasma sources, and available diagnostic techniques to provide a starting point for advanced research into this field.
\end{abstract}


\maketitle
\thispagestyle{ARTTITLE}

\section{Introduction}

The properties of the plasma crucially control the attributes of the accelerated particle beam in beam-driven (PWFA) and laser-driven (LWFA) plasma wakefield accelerators. Tailoring the characteristics of plasma is a powerful tool to control beam injection, the transverse focusing forces, and the acceleration gradient acting on a witness beam. In addition, plasma features can govern the evolution of the driving laser or driving particle beam exciting the wakefield. Furthermore, engineered plasmas allow for manipulation of beam phase space in active plasma lenses and dechirpers. Therefore, accurate control over the~plasma density and temperature profile is a prerequisite to realise design beam specifications.

\section{Plasma generation mechanisms}

Plasma generation is fundamentally concerned with the ionisation of a background medium, usually a~gas, to produce electrons and ions in a controlled way. 

\subsection{Electrical discharges}

The most important collisional ionisation mechanism is electron-impact ionisation, whereby during a~collision an electron of sufficient energy overcomes the ionisation potential of an electron bound to an atomic core to create an electron-ion pair.  

\begin{figure}[t]
    \centering
    \includegraphics[width=0.55\linewidth]{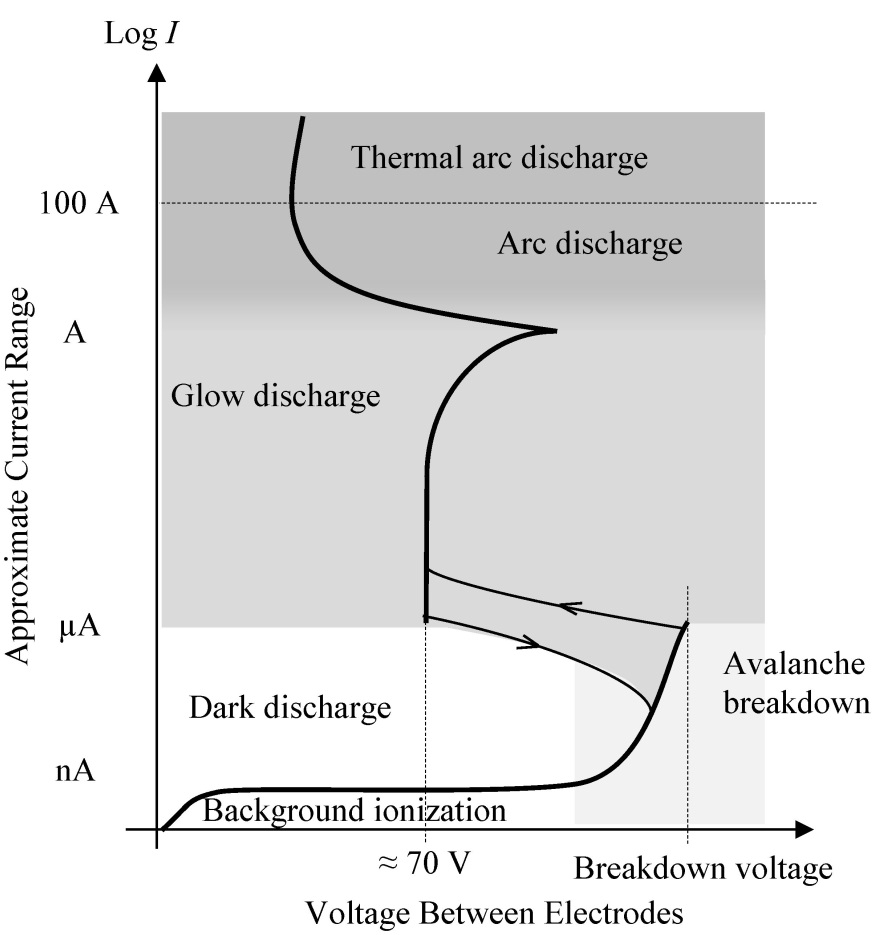}
    \caption{The current–voltage characteristic of a typical electrical discharge between two electrodes. Image from Ref. \cite{Fair2014} (CC BY 4.0).}
    \label{fig:VIdiagram}
\end{figure}

In an electrical discharge, free electrons gain energy from an externally applied electric field.~There are many basic types of phenomena that comprise electrical discharges, including electron avalanches, streamers, leaders, glows and arcs. The region where some common discharge types exist in voltage-current space is shown in Fig. \ref{fig:VIdiagram}.
For brevity we will focus here on the Townsend avalanche mechanism. For more discussion on other electrical discharge mechanisms, see Refs. \cite{KunLue2013,LiebLich2005}. A voltage applied between two electrodes causes the charge carriers to gain energy. There is always some low level of background ionisation due to background radiation, e.g. from cosmic rays, which may trigger the avalanche process. Above a critical field, the existing free electrons gain enough energy to ionise the background neutrals. These secondary ionised electrons (and the now-slowed primary electrons) both gain energy until they can also ionise, and so on, causing an exponential growth in the free electron density, $n_e$. This exponential electron density increase is known as a Townsend avalanche, and can be characterised by the~following rate equation (in one dimension, $x$):
\begin{align}
    \frac{dn_e}{dx} &= \alpha n_e,
\end{align}
which has the solution $n_e(x) = n_{e}(0)\exp\left(\alpha x\right)$. The coefficient $\alpha$ is called the primary ionisation coefficient, or the first Townsend coefficient. 

\begin{figure}[ht!]
    \centering
    \includegraphics[width=0.8\columnwidth]{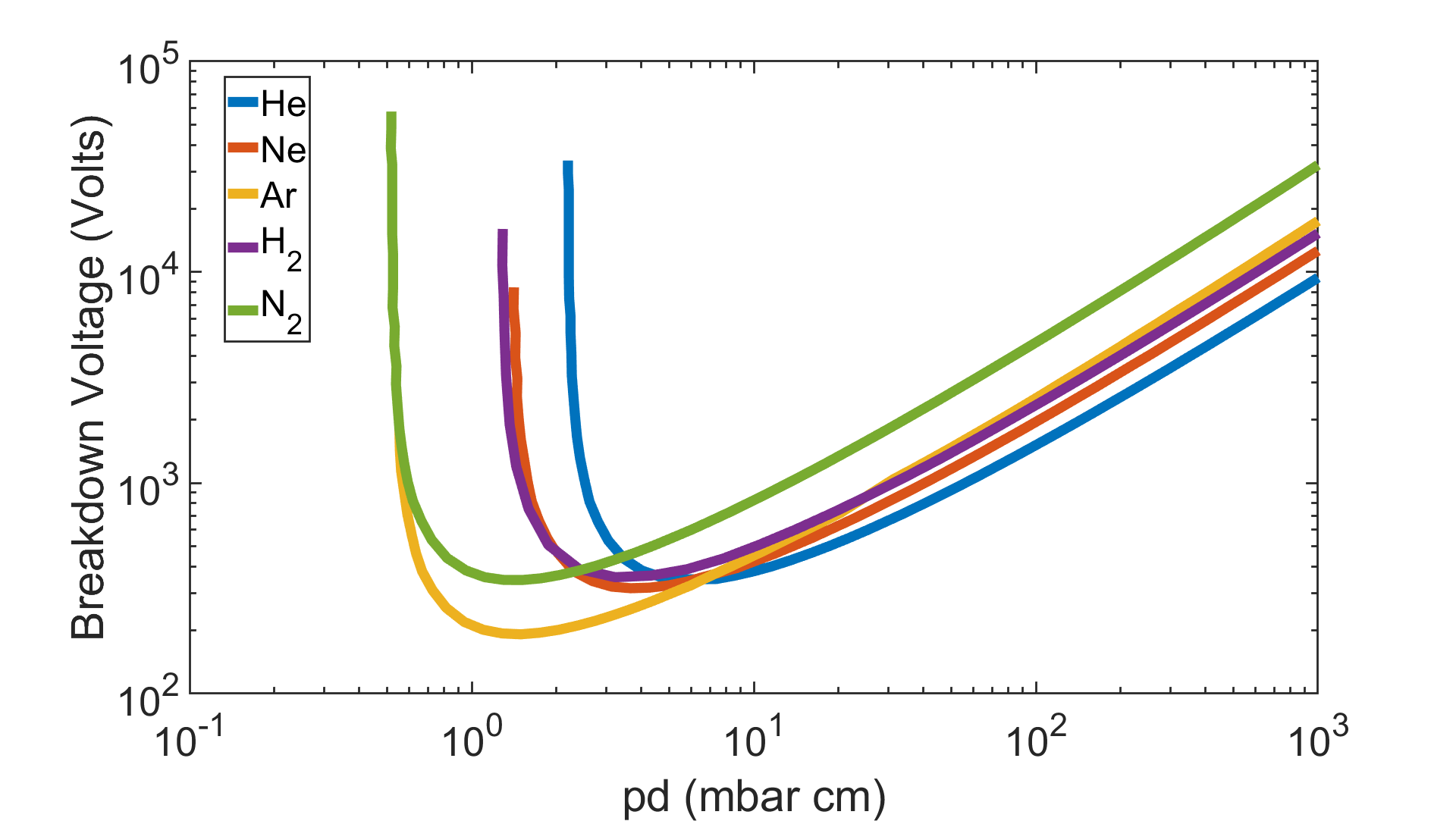}
    \caption{Paschen curves for helium, neon, argon, molecular hydrogen and molecular nitrogen.}
    \label{fig:Paschen}
\end{figure}

The breakdown voltage (or critical electric field) of a gas between two flat electrodes fundamentally depends on the electron mean free path (the average distance between electron-atom collisions) and the distance between the electrodes. The electron mean free path $\lambda_{MFP}$ is directly related to the inverse of the gas pressure $p$, i.e.,
\begin{align}
    \lambda_{MFP} = \frac{1}{\sigma n} = \frac{k_bT}{\sigma p},
\end{align}
where $\sigma$ is the effective cross-sectional area for collisions, $k_b$ is Boltzmann's constant, and the ideal gas relation has been used to connect the gas density to pressure. At very low pressures, when the~electron mean free path is longer than the distance between the electrodes, the electrons are unlikely to undergo any collisions before they reach the anode, and hence a very large voltage is required to start the discharge. Conversely, at very high pressures the electron mean free path is very small and electrons do not get enough time to accelerate to ionising energies before collisions with atoms reduce their energy once again. Between these two extremes there exists a minimum breakdown voltage that depends on gas type and electrode material, i.e., a balance between the free acceleration time and likelihood of collisions. The~relation between the product of the pressure and distance between electrodes, and the voltage yields the well known Paschen curve (see Fig. \ref{fig:Paschen}).
Avalanches and streamer breakdowns operate in the~region to the right of the minimum. The region to the left of the Paschen minimum represents vacuum breakdowns, where the voltage required to initiate breakdown increases rapidly by several orders of magnitude. The~electrode surface plays a critical role in vacuum breakdowns, as the charge carriers required to create a conductive plasma channel between the electrodes now originate from the electrodes themselves.

\subsection{Laser Ionisation}
\label{para:lasion}
A convenient method of ionising material is to use a laser.
Single photon ionisation is possible when the~photon energy $\hbar \omega$ ($\hbar$ is Planck's constant, $\omega$ is the angular frequency) is greater than the binding potential of the atom $\mathcal{E}_{\mathrm{ion}}$. However, for Hydrogen $\mathcal{E}_{\mathrm{ion}} = 13.6$\,eV, which is significantly higher than single photon energies of commonly used optical/near-infrared lasers (\SI{800}{\nano\metre} photons have $\hbar \omega = 1.55$\,eV).
If the laser intensity is high enough then there can be a large enough number of photons in the atomic volume to give a high probability of multiple $(m)$ photons simultaneously ionising the atom, with $m \hbar \omega > \mathcal{E}_{\mathrm{ion}}$.
This process is called multi-photon ionisation. More information on this, and on all of the mechanisms discussed here, can be found in Ref. \cite{Gibbon}.

An alternative route to ionisation is barrier suppression ionisation, where the laser provides an~electric field greater than that felt by the valence electron from the (shielded) nucleus.
If, in 1D, we apply an~electric field $E$ to an atom, the potential as a function of distance from the nucleus $x$ is given by
\begin{equation}
    V(x) = -\frac{Ze^2}{4\pi \epsilon_0 x} - eEx,
\end{equation}
where $Ze$ is the charge of the atom, minus the electron under consideration. Here $e$ is the elementary charge, and $\epsilon_0$ is the vacuum permittivity. 
It can be shown that the position of the potential maximum is $x_{\mathrm{max}} = \sqrt{Ze/4\pi \epsilon_0 E}$.
By setting $V(x_{\mathrm{max}}) = \mathcal{E}_{\mathrm{ion}}$ a threshold electric field, $E_t$, above which barrier suppression ionisation occurs is
\begin{equation}
    E_t = \frac{\pi \epsilon_0 {\mathcal{E}_{\mathrm{ion}}}^2}{Ze^3}.
    \label{eq:Ethreshold}
\end{equation}
Written as a threshold intensity, $I_t$, this is
\begin{equation}
    I_t = \frac{\pi^2 {\epsilon_0}^3 c}{2e^6} \frac{{\mathcal{E}_{\mathrm{ion}}}^4}{Z^2} \approx 4 \times 10^9 \left(\frac{\mathcal{E}_{\mathrm{ion}}}{\mathrm{eV}} \right)^4 \frac{1}{Z^2} \ \si{\W\per\square\centi\metre}.
\end{equation}
Specifically, $\mathcal{E}_{\mathrm{ion}}$ is the ionisation potential of an atom with charge $(Z-1)e$.
As examples, this equation predicts that Hydrogen will be ionised at \SI{1.4e14}{\W\per\square\centi\metre} and Helium will be fully ionised at \SI{8.8e15}{\W\per\square\centi\metre}.
A consequence of this is that pre-ionised plasmas are not necessarily required for LWFA studies, where the leading edge of the $10^{18}-10^{19}$\,\si{\W\per\square\centi\metre} pulse is sufficient to create a plasma.\\
At electric fields approaching $E_t$ the atomic potential is already sufficiently distorted such that an electron has a chance of tunnelling out of the potential. 
The probability of this must be found from the~solution to the Schr{\"o}dinger equation.
Simplified models of this exist.
A popular one, often employed in particle-in-cell codes, is the ADK (Ammosov, Delone and Krainov) model \cite{ADK}.
Some examples of models that are valid to higher field strengths, although they assume static fields, include Refs. \cite{Tong_2005,Zhang2014}. 
The Keldysh parameter $\gamma_K = \omega \sqrt{2\mathcal{E}_{\mathrm{ion}}/I}$, where $I$ is the laser intensity, distinguishes between the multiphoton $(\gamma_K>1)$ and tunnelling $(\gamma_K<1)$ regimes.

\subsection{Beam Ionisation}
High peak current, tightly focused electron beams can have Coulomb fields that exceed the threshold electric field for barrier suppression ionisation from Eq. (\ref{eq:Ethreshold}).
We start by considering the simplest case of a uniform cylinder of electrons of number density $n_b$ and radius $r_b$.
Applying Gauss's law to this system allows us to determine the electric field inside and outside the cylinder as a function of radial distance $r$ to be
\begin{equation}
    E_{\mathrm{in}} = -\frac{en_b r}{2\epsilon_0},
\end{equation}
\begin{equation}
    E_{\mathrm{out}} = -\frac{en_b{r_b}^2}{2\epsilon_0}\frac{1}{r}.
\end{equation}
Note that the maximum electric field is not on-axis but at $r=r_b$.
For example, such a beam with density $n_b = \SI{1e18}{\per\cubic\centi\metre}$ and $r_b = \SI{5}{\um}$ has a maximum electric field of \SI{45}{\giga\volt\per\metre} compared to $E_t = \SI{32}{\giga\volt\per\metre}$ for Hydrogen.
For a gaussian beam the electric field varies as $E_r \propto \left[1 - \exp(-R^2/2) \right]/R$, where $R = r/\sigma_r$. This has a maximum at $r\approx 1.6 \sigma_r$.
However, beam driven wakefield experiments tend to utilise a pre-ionised plasma to combat head erosion, where the head of the beam is free to diverge because it is propagating through neutral gas and thus not situated inside the focusing phase of the~wakefield.

\section{Tailoring plasma properties to control wakefield and plasma processes}

Plasma sources exist in a number of variants. Common technologies are gas jets, gas cells, Alkali vapour ovens, or discharge waveguides, all of which are discussed in the following. In general, these sources are designed to provide plasma created by ionising a neutral gas species. The geometry of the gas delivery and the method by which the gas is turned into plasma crucially affect the conditions of the resulting medium for acceleration.

The geometry of the source effectively determines the gas density distribution in space and time. The means of gas ionisation, either by a laser, an electrical discharge, or the field of a high-current density particle beam, control the plasma density and temperature distribution. The temperature evolution can additionally be affected by the material composition of the source and its thermal properties. This interaction can, for example, lead to the creation of a laser guiding channel. Such plasma sources with guiding capabilities typically consist of a cylindrical channel cut from a hardened substance, usually sapphire, into which a gas is fed and ignited by a discharge. The choice of material in this case is critical for controlling the heat flow, but also for the durability of the construction. Many other aspects must be considered when designing such and similar plasma sources. These are for example:
\begin{itemize}
    \item The length of the source defines the maximum plasma column length and hence influences the~total energy gain in the wakefield and the total efficiency of the process.
    \item The diameter of the plasma column can facilitate the guiding of a laser in plasma, which may be affected by the radial expansion rate of the plasma column and its temperature gradients.
    \item The gas system may be utilised to feed a variable flow of neutral gas into the source vessel as the~gas density controls the plasma density.
    \item The gas species should be carefully chosen as gases with different ionisation potentials and electronic levels determine the created plasma density and influence wakefield, laser, and particle beam dynamics.
    \item The detailed geometry of the gas inlets and outlets is of importance to tailor the initial gas profile before ionisation. A density gradient on the exit of a plasma source may be important to control the beam release.
    \item The material from which the vessel is constructed defines the durability under heat load from plasma and/or the laser interaction. Ultimately, this may limit the repetition rate and lifetime of the source.
\end{itemize}

\subsection{Plasma density considerations} 
When considering the design of a wakefield acceleration stage the chief constraint is usually the laser peak power, or in the case of beam driven wakefields, the peak current. 
Once the driver conditions have been specified, sensible estimates of the optimum plasma density $n_e$ can be made.

For laser wakefield accelerators (LWFAs) in the absence of a laser guiding structure (Section \ref{sec:plasmawaveguides}) a minimum $n_e$ is set by the self-guiding criterion (Section \ref{para:laserwaveguides}).
In the laser and beam driven cases driving a high amplitude wakefield requires that the length of the driver is shorter than the wakefield period i.e. the laser pulse length $c\tau_0 \leq \lambda_p/2$ (for linear LWFAs this drives the wakefield resonantly) or the beam length $\sigma_z = \sqrt{2}/k_p$ (for beams with $\sigma_r \ll \sigma_z$), where $k_p$ and $\lambda_p$ are the plasma wavenumber and wavelength \cite{Lu2005, Lu2010}.
Both of these conditions can provide an `optimum' density.
Laser/ particle beam focusing is then determined by this density.
In LWFAs the laser will maintain a constant width and intensity, and thus drive a constant amplitude wakefield, if the matched spot size condition $k_p w_0 =2\sqrt{a_0}$ is met \cite{Lu2007}.
Most experiments are performed with a normalised vector potential $a_0 \approx 1-4$ to access the high accelerating fields of the blowout regime.
For beam driven wakefields the beam will maintain a constant width, and thus drive a constant amplitude wakefield, if its divergence is cancelled out by the~focusing force of the wakefield.
This occurs when $\sigma_r = \sqrt{\epsilon/k_{\beta}}$, where $\epsilon$ is the emittance of the~beam (see Section \ref{para:ramps}) and $k_{\beta} = k_p/\sqrt{2\gamma}$ is the betatron wavenumber \cite{Litos2019}.

Above we have described the common process by which a sensible value of $n_e$ can be obtained. 
However, these criteria only set the scale for the plasma density. 
$n_e$ also features in the trapping condition, so is a crucial parameter for any injection mechanism. 
It may be increased or decreased to promote/suppress self-injection. 
In a plasma accelerator the complete properties of the driver may not be known (e.g.~unknown laser wavefront or spatio-temporal couplings or missing high precision measurements of the 6D phase space of the drive beam). 
Additionally the relationship between gas backing pressure and peak density and the density distribution, which may be longitudinally nonuniform, may not be precisely known, especially for complex plasma sources. 
Thus, it is imperative to be able to change the plasma density precisely in a density range determined from the optimum parameters.
This can be done by changing the backing pressure of the gas, or where an ionisation laser or discharge is used to create the plasma, by changing the timing between the the ionising pulse and the wakefield drive pulse, as the plasma density reduces over time due to recombination and plasma expansion.\\

\subsection{Plasma length limitations} 
For LWFAs the ideal total length of the plasma is usually given by the pump depletion length $L_{pd} = \frac{{\omega_0}^2}{{\omega_p}^2} \frac{\omega_p}{k_p} \tau$, or the dephasing length $L_d = \frac{4}{3} \frac{{\omega_0}^2}{{\omega_p}^2} \sqrt{a_0} {k_p}^{-1}$ \cite{Lu2007}.
The latter is often chosen as we wish to extract the electrons from the plasma when their energy is maximal. 
However, the precise length we require can be longer than $L_d$ to allow for laser evolution before beam injection occurs. 
For externally controlled injection methods such as external beam injection or two-colour ionisation injection \cite{Bourgeois2013}, for example, the positioning of the injected beam may not allow for acceleration along a full dephasing length, so the optimal length will be shorter. 
Thus it is also desirable to have an adjustable length plasma source. 
Such a source can be used to study electron acceleration as a function of plasma length, which can be used to infer the accelerating fields of the wakefield \cite{poderthesis}.
It is generally not the case that, for a~fixed length source, the optimal density for stable guiding of the drive beam is the optimal density for maximum electron energy gain.
An adjustable length source allows these two effects to be decoupled.
An example of an adjustable length gas cell can be found in Section \ref{sec:gascells}.
Other works have decoupled the injection and acceleration processes using a high density, short injector stage followed by a longer, lower density acceleration stage \cite{Pollock2011,Kim2013}.

PWFAs do not suffer from a similar dephasing effect (except via beam head erosion). 
The maximum length of the accelerator is limited by the distance over which the drive beam is decelerated by the fields of the wakefield it is driving. 
There is, therefore, a density dependence in the ideal plasma length.
This length also depends on the characteristics of the wakefield, which are determined by the radial and longitudinal bunch dimensions as well its charge \cite{Rosenzweig2004}, and so the optimum length is often found by simulation, e.g. Ref.~\cite{Lotov2007}.
Therefore, future PWFAs could benefit from adjustable length plasma sources.

\subsection{Laser guiding}
\label{para:laserwaveguides}
In order to reach the intensities required to drive a nonlinear wakefield with a laser, tight focusing is used.
For example, a 1\,J, 30\,fs FWHM, $\lambda = 800$\,nm laser pulse must be focused to a spot size $w_0 = $ \SI{15.3}{\um} to reach $a_0 = 2$, where $w_0$ is the $1/e^2$ intensity width. 
The Rayleigh range $z_R = \pi {w_0}^2 / \lambda$, the distance over which the intensity halves due to diffraction of a Gaussian focused beam, is \SI{0.92}{\milli\metre} for this focus.
This compares unfavourably with the dephasing length for the matched density, which is $L_d \approx 18$\,mm.
Indeed it is often the case that $L_d > z_R$.
Therefore, strategies are required to guide the laser pulse at a~constant radial width over the full accelerator length so that a constant, large amplitude wakefield can be driven.

For a ${a_0}^2/4 \ll 1$ and ${\omega_p}^2 \ll {\omega_0}^2$, the nonlinear refractive index of the plasma can be written as
\begin{equation}
    \eta \approx 1 - \frac{1}{2} \frac{{\omega_{p0}}^2}{{\omega_0}^2} \left( 1 + \frac{\delta n_e}{n_{e0}} - \frac{2 \delta \omega}{\omega_0} - \frac{{a_0}^2}{4} \right),
    \label{eq:refractiveindex}
\end{equation}
where $\omega_0$ is the central laser frequency and $n_{e0}$ is the background plasma density with associated plasma frequency $\omega_{p0}$ \cite{Mori1997}.
The ${a_0}^2/4$ term is due to the relativistic mass increase of the plasma electrons in the laser field.
Since a Gaussian laser pulse has an intensity maximum on axis, the plasma frequency is lowest there, corresponding to a maximum of the refractive index.
A radially decreasing refractive index, with a maximum on axis, is analogous to a lens.
Thus the plasma can be used to focus the beam.
By balancing diffraction and the focusing effect of this plasma lens \cite{Sprangle1987}, it can be shown that a laser will be self-guided for powers greater than the critical power
\begin{equation}
    P_c \approx 17 \frac{{\omega_0}^2}{{\omega_p}^2} \ \mathrm{GW}.
    \label{eq:Pc}
\end{equation}
Self-guiding has been shown to occur over as many as 100 Rayleigh lengths \cite{Poder_2017}, and there are multiple GeV level LWFA experiments that have been operated in a self-guided regime \cite{Kneip2009,Clayton2010,Kim2013,Wang2013}.

Self-guiding is the simplest method, but it has been shown that electron energy gain is lower in this regime than when a similar power pulse is guided by an externally imposed radial plasma density profile \cite{Leemans2014,Gonsalves2019}.
A Gaussian laser pulse can be guided in a parabolic density profile of the form $n_e(r) = n_e(r=0) + \Delta n_e r^2/{r_{ch}}^2$. 
Such a laser pulse will be guided in this plasma profile by a similar plasma lens effect, at a spot size of $w_m = \left({r_{ch}}^2/ \pi r_e \Delta n_e \right)^{1/4}$, where $r_e$ is the classical electron radius \cite{Esarey1997}.
This neglects ponderomotive and relativistic effects.
A parabolic density profile can be produced in a capillary discharge as described in Section \ref{para:dischargewaveguides} and in Refs. \cite{Spence2000,Butler2002}, by laser heating of a column of a collisional plasma \cite{Durfee1993}  as described in Section \ref{para:hydroewaveguides} or by optical field ionisation of a lower density plasma \cite{Lemos2013,Shalloo2019}.
The amplitude of the refractive index change $\Delta n_e$ can be increased by further laser heating so that smaller laser spots can be guided in a given plasma density $n_e$ produced by the discharge \cite{Gonsalves2019}.

\subsection{Injection from plasma structures}
In order for an electron to be (internally) injected into a wakefield it must fulfil a trapping condition.
In the frame moving at the wakefield phase velocity, a particle will be trapped in the wakefield if its kinetic energy $(\gamma_e' - 1)m_ec^2$ is less than the potential of the wakefield $e\phi'$.
Lorentz transforming to the lab frame, this condition becomes
\begin{equation}
    e\phi > \left[ \gamma_e \left( 1-\beta_e \beta_{ph} \right) - \frac{1}{\gamma_{ph}}\right] m_ec^2,
    \label{eq:trapping}
\end{equation}
where $\beta_{ph}c$ is the wakefield phase velocity and $\gamma_{ph} = \omega/\omega_p = \left(1 - \beta_{ph} \right)^{1/2}$.
A more thorough discussion of trapping can be found in \cite{Esarey1995}.
For a cold background plasma $(\beta_e=0)$ with large $\gamma_{ph}$ this condition becomes $e\phi \gtrsim (1-1/\gamma_{ph})m_ec^2$.
This can be re-written as $E \gtrsim E_0 = m_ec\omega_p/e$, i.e.~the cold non-relativistic wavebreaking limit, a coincidence which is often used to explain self-injection.\\
One method to promote injection more easily, and in a controlled manner, is to use a density downramp in the plasma density profile \cite{Bulanov1998,Geddes2008}.
This causes $\gamma_{ph}$ at the back of the wakefield bubble to reduce rapidly as the plasma wavelength increases with decreasing density, significantly lowering the trapping threshold throughout the density gradient.
It can be shown \cite{DeLaOssa2017} that $\beta_{ph}$ can be written as
\begin{equation}
    \beta_{ph} = \beta_d \left(1 + \frac{\chi}{2\Tilde{n}^{3/2}} \frac{1}{k_{p0}} \frac{\mathrm{d} \Tilde{n}}{\mathrm{d}z} \right)^{-1},
    \label{eq:vgramp}
\end{equation}
where $\beta_d c$ is the driver velocity, $\Tilde{n} = n/n_0$ is the density relative to the background density, $k_{p0}$ is the~plasma wavenumber associated with $n_0$ and $\chi = k_p(z-\beta_d ct)$ is the phase of an electron in a plasma wave, which is increasingly negative behind the driver.
Here we explicitly see that in a negative gradient $\mathrm{d}\Tilde{n}/\mathrm{d}z < 0$, electrons near the back of the plasma wave (large negative $\chi$) experience a much lower $\beta_{ph}$ and can be easily trapped.
A density downramp can be produced in a two compartment gas cell, such as the example in \cite{KONONENKO2016125}, by additional localised laser ionisation \cite{WITTIG201683} or by having two gas jets close together.
Injection can also occur in the downramp at the end of the gas sources, which is how it was originally discovered \cite{Geddes2008}, but can be an unwanted source of dark current.

A more extreme version of this is shock injection, where a razor blade is introduced into a supersonic gas flow target (see `Gas Jets', Section \ref{sec:GasJets}) to produce a shock, resulting in a discontinuity in the~plasma density.
This causes the plasma wavelength to greatly increase over a short distance, trapping a significant number of electrons from the sheath in a localised region of the wakefield \cite{Suk2001}.
As such, this method has been proven to produce high charge, low energy spread beams in a controllable manner, e.g.~Ref. \cite{Buck2013}.\\

\subsection{Density ramps} 
\label{para:ramps}
Once electrons have been injected into a wakefield they are rapidly accelerated to a velocity close to $c$, while in a constant-density LWFA $\beta_{ph} = v_g/c < 1$, and the electrons dephase.
It is, however, possible to take inspiration from Eq. (\ref{eq:vgramp}) and construct a density upramp such that, at the back of the wakefield, $\beta_{ph} \approx c$.
In this way the electron beam can be fixed in a constant accelerating phase and not suffer from dephasing, while also suppressing additional unwanted injections.
Other limitations such as pulse depletion still occur, but enhanced energy gains are achievable in this and other schemes \cite{Rittershofer2010,Dopp2016}.

Now we have exploited density gradients for high quality injection and enhanced acceleration, we can turn to them once again to ensure a high quality beam is extracted from the wakefield accelerator. 
By high quality we mean low normalised emittance $\epsilon_n = \left\langle \gamma \beta \right\rangle \sqrt{\left\langle x^2 \right\rangle \left\langle x'^2 \right\rangle - \left\langle xx' \right\rangle^2} = \left\langle \gamma \beta \right\rangle \epsilon$, where $x$ and $x' =  p_x/p_z$ are the coordinates of the~beam electrons in the transverse phase space.
It can be shown that a beam with finite energy spread $\sigma_E$ and divergence $\sigma_{x'}$ experiences transverse emittance growth in free space drift, distance $s$, according to the equation ${\epsilon_n}^2 \approx \left\langle \gamma \beta \right\rangle^2 \left( {\sigma_E}^2 {\sigma_{x'}}^4 s^2 + \epsilon^2\right)$ \cite{Floettmann2003}.
For large energy spread beams (per-cent level), this equation is extremely prohibitive for the transportation of large divergence beams (milliradian divergence).
This emittance growth can be mitigated by a density downramp at the end of the plasma stage, where the focusing forces of the wakefield are adiabatically reduced, allowing the beam size to increase under gradually reducing focusing forces, decreasing the~beam divergence at the plasma-to-vacuum interface and thus the emittance growth rate \cite{MEHRLING2016367}.

In a blown-out plasma wakefield the electron beam is matched to the wakefield if the beta function of the beam $\beta \approx c/\omega_{\beta}$.
An externally injected witness beam must be matched to preserve normalised emittance throughout its acceleration.
The small plasma length scale $c/\omega_{\beta}$ means that small (millimetre) $\beta$ functions are required, which can be difficult to achieve.
In this case a matching section consisting of a density upramp at the start of the plasma can be used to reduce emittance growth in the case of initially unmatched witness beams.

\begin{figure}[t]
    \centering
    \includegraphics[width=0.95\linewidth]{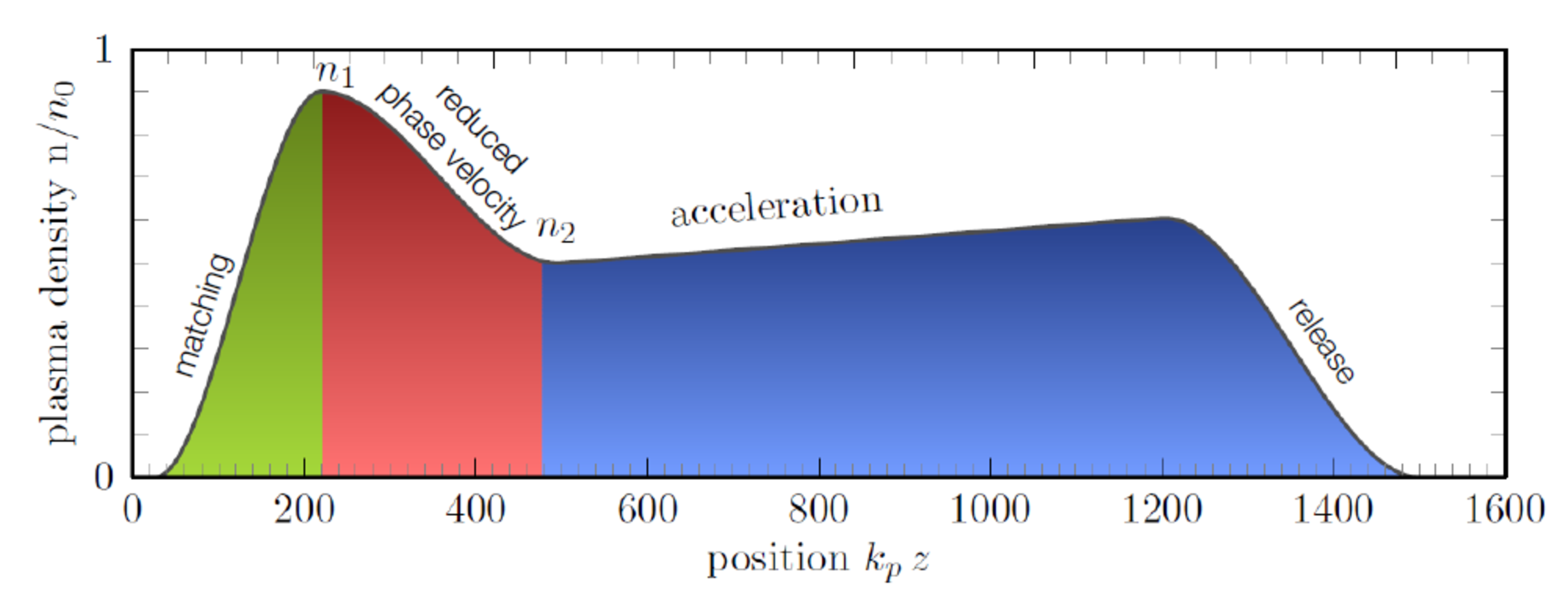}
    \caption{A tailored plasma stage with input and output matching, trapping, and dephasing-beating acceleration stages.}
    \label{fig:plasstage}
\end{figure}

Combining everything we have learned so far, an ideal plasma source would be based on Fig.~\ref{fig:plasstage}.
Even more advanced schemes can be considered for the optimisation of the electron beam quality by tailoring the longitudinal plasma density.
For example, tapered density transitions can be used to help mitigate the hosing instability in beam driven wakefield accelerators \cite{Mehrling2017}.
In a plasma accelerator the~accelerating field varies along the bunch length which produces a large energy chirp.
In the absence of beam loading a modulated plasma density can be used to significantly reduce chirp \cite{Brinkmann2017}.

\subsection{Plasma dopants for injection} 
An alternative to self-injection or controlling injection with the plasma density is to control injection via the plasma composition.
In ionisation injection the background plasma (typically Hydrogen or Helium) is doped at the per-cent level with a higher atomic number gas.
The gas is chosen such that its $k$-shell electrons are ionised via barrier suppression ionisation (Section \ref{para:lasion}) only at the peak of the laser pulse, and electrons are `born' inside the wakefield cavity \cite{Umstadter1996}.
A common dopant is Nitrogen.
This is considered to be a robust injection technique, which may produce higher charge and higher quality beams than self-injection \cite{McGuffey2010,Pak2010}.
However, the ionisation, and thus the injection, is a continuous process and may produce large energy spread beams.
Lower bandwidth beams can be produced via self-truncated ionisation injection \cite{STII}, where the injection happens only over a short distance due to over-focusing of the laser pulse changing the wakefield potential.
Low bandwidth, short pulse, low emittance beams (i.e.~beams with high 6D brightness) can be produced via localised ionisation injection occurring in a~small 3D volume initiated by a tightly focused, additional laser pulse in a LWFA \cite{Bourgeois2013}, or in a PWFA in the so-called Trojan Horse scheme \cite{TrojanHorse}.

\subsection{Plasma temperature effects}
\label{para:temperature}

 The average kinetic energy of an electron subject to the ponderomotive force is $\approx 100$ keV for $a_0 = 1$ and a laser wavelength of 1 $\mu$m. This is many orders of magnitude larger than typical plasma temperatures of $<10$ eV, and hence temperature effects are most often neglected when describing the formation, evolution and breaking of wakefields.  However, the plasma component temperatures (and densities) are of high importance for non-wakefield applications, e.g.~laser waveguides and active plasma lenses.

On large enough time and space scales the plasma components (electrons, ions and neutral species) behave like (electromagnetic) fluids, and can be described by transport equations and transport properties akin to more common fluid flows. For example, the ideal magnetohydrodynamics \cite{hosking2016fundamental} model describes plasmas by continuity and momentum equations,
\begin{align}
    \frac{\partial \rho}{\partial t} + \nabla\cdot\left(\rho v\right)&= 0, \\
    \rho\left(\frac{\partial v}{\partial t} + v\cdot\nabla\right)v &= J\times B - \nabla p
\end{align}
which are coupled to Maxwell's equations of electromagnetism and an equation of state. Here $\rho, v$ and $p$ are the plasma bulk mass density, velocity and pressure, respectively. $J$ is the current density and $B$ is the~magnetic field, which may be applied externally or induced internally. In a plasma, transport properties such the electrical resistivity, thermal conductivity and collision rates generally depend directly on the temperature and density of all components.  

In a discharge-formed plasma, the discharge current causes resistive heating. Thermal conduction and fluid convection allows for the re-distribution of thermal energy throughout the plasma, and thermal transfer with the plasma boundary represents energy loss. Competition between resistive heating and boundary heat loss dominate the plasma evolution. For a cylindrical plasma capillary a temperature gradient forms between the `hot' capillary axis and the `cool' wall. This gradient is exacerbated by the~temperature dependence of the electrical resistivity, $\eta \propto T^{-3/2}$, which results in a decreased resistance in the hotter on-axis region, leading to preferential current flow and even greater temperatures. The~temperature gradient creates a pressure gradient, which forces plasma material towards the wall and leads to a density minimum on axis. In equilibrium, where the heating due to the current is perfectly balanced by the heat loss to the wall, it can be shown that the electron density will form a near-parabolic profile radially which can be exploited for laser waveguiding (further discussed in Section \ref{para:dischargewaveguides}). In contrast, the temperature and density variation can be detrimental to the operation of active plasma lenses (further discussed in Section \ref{para:activeplasmalens}).

\subsection{Ion mass effects}

For the majority of plasma applications under consideration it is the electron density that is the most important property, and the background medium from which the plasma/ions are created is often considered to be of negligible importance. However, the plasma neutral and ionic species (collectively called the heavy species) can have critical influences.

The most obvious effect of the ion mass is the resistance to momentum changes. The light electrons respond to perturbations, such as those caused by a drive beam in PWFA or a high powered laser in LWFA, on very small timescales (fs$-$ps). The neutral/ion masses are $>10^3$ times larger than the~electron mass, such that the response time is also much larger. The plasma frequencies for electrons, $\omega_e$, and ions, $\omega_i$, is given by
\begin{align}
    \omega_e &= \sqrt{\frac{n_e e^2}{m_e\epsilon_0}}, \\
    \omega_i &= \sqrt{\frac{n_i Z^2 e^2}{m_i\epsilon_0}},
\end{align}
where $n_e$ and $n_i$ are the electron and ion densities, $m_e$ and $m_i$ are the electron and ion masses, and $Z$ is the mean ion charge. Thus from the perspective of the electrons responding to perturbations, the ions appear static. For this reason, modelling phenomena on the electron response scale (i.e. PWFA) can often ignore ion motion which vastly simplifies calculations and computational time/effort. However, at ps$-$ns time scales the ion motion becomes important, e.g. with onset of phase-mixing \cite{SenGupta1999}, expansion of the plasma column \cite{Gilljohann2019}, and significant ion-impact collisional ionisation. Now simulations cannot ignore the ion motion leading to more complex calculations. The dampening of these oscillations (and return to an unperturbed state) is also directly related to the ion mass, with implications to high repetition rate operation.

\section{Technical implementation of plasma sources}

\subsection{Gas jets}
\label{sec:GasJets}
\begin{figure}[t]
    \centering
    \includegraphics[width=0.8\columnwidth]{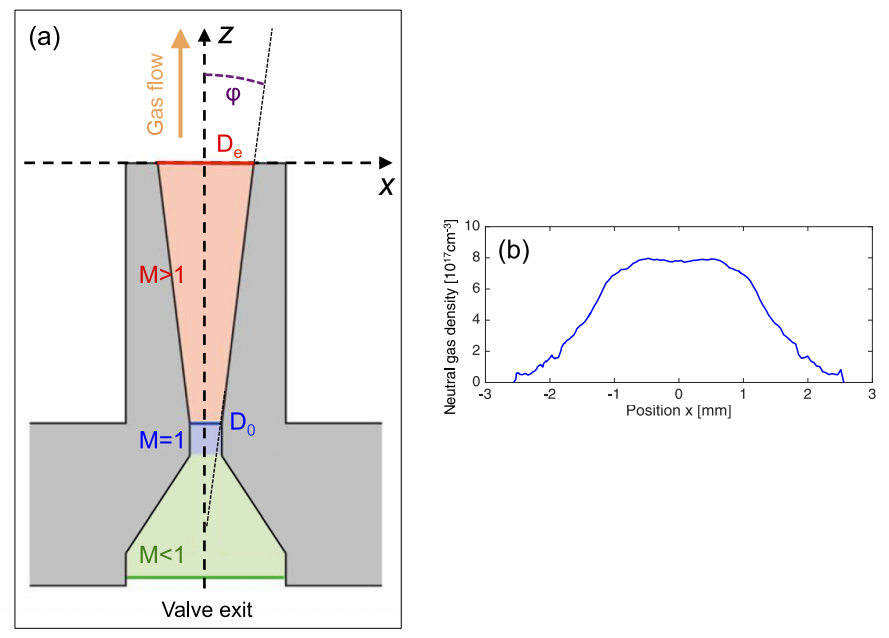}
    \caption{An example of a supersonic gas jet. (a) A fast valve releases a pulse of gas from a backing reservoir through a constriction and in to a conical nozzle. The jet is tailored by choice of inner diameter $D_0$, exit diameter $D_e$ and opening angle $\psi$. $M$ represents the Mach number in each region. (b) Optimising the nozzle parameters produces a flat top density profile above the nozzle, as illustrated. Image from Ref. \cite{Semushin2001} (CC BY 4.0).}
    \label{fig:gasjet}
\end{figure}

The lowest order approximation to an ideal density profile is a flat-top, which allows a constant amplitude wakefield to be driven throughout the plasma length (ignoring driver evolution).
A near-flat-top profile can be produced in a supersonic gas jet.
As depicted in Fig. \ref{fig:gasjet}~(a), high pressure (multiple bar to 100 bar) is held in a backing volume and is released in to vacuum via a fast valve, where it is forced through a constriction and then allowed to expand in a conical nozzle. 
This nozzle design is an approximation of the well known de Laval nozzle.
Figure \ref{fig:gasjet}~(b) is an example of the gas density profile above the nozzle.
To create the desired density profile, the height of the nozzle must be optimised with respect to the throat and exit diameters, as discussed in \cite{Semushin2001}.

Supersonic gas nozzles have been favoured in many works because of their simplicity, 360 degree diagnostic access and the ability to create shock fronts for shock injection.
The main disadvantage is the large volumes of gas they eject, even when operated in pulsed mode, which is prohibitive for high repetition rate operation.
This can be partly mitigated by using a slit nozzle or by additional high capacity pumping.
A slit nozzle has inlets and outlets that are asymmetrical; they are much longer in the drive laser direction than the transverse direction, e.g. Ref.~\cite{COUPERUS2016}.
As such they reduce gas load compared to cylindrical nozzles.
Additional disadvantages of gas jets include the erosion of mechanical parts and vibrations caused by the release of high pressure gas and the mechanical motion of valve parts.
Ripples in the gas density profile, which can be caused by imperfections on the nozzle surface, can caused unwanted self-injection and acceleration of dark current \cite{Kushcel2018}. 

\subsection{Gas cells}
\label{sec:gascells}
Another common source type is the gas cell.
Effectively, this is a box with a pinhole at each end to allow the driver to enter and exit the cell.
The cell is filled with gas in the few millibar to 1 bar range to give molecular densities in the range $10^{17} - 10^{19}$\,\si{\per\cubic\centi\metre}.
This significantly reduces the gas load to the chamber compared to gas jets, allowing gas cells to be run in continuous flow mode with additional pumping.
They have the additional advantage that they have no moving parts, contributing to the improved shot-to-shot stability of electron beams from gas cells compared to gas jets \cite{Osterhoff2008}.
Gas cells may contain a~uniform gas distribution with density downramps going through the pinholes, with ramp scale lengths of approximately the pinhole diameter \cite{KONONENKO2016125}.
Like gas jets, laser ionisation is required (this can be the~front of an intense drive laser). 
The pinholes should be sufficiently large as to reduce damage from higher order modes in the focal spot, as their erosion changes both the gas density and profile.
Multiple designs of adjustable length gas cells exist.
Diagnostic access is reduced somewhat compared to gas jets, however cells are often made with windows for transverse laser probing.

\subsection{Heat-pipe ovens}
To create metre scale plasmas at low density ($10^{14}-10^{17}$\,\si{\per\cubic\centi\metre}) with high uniformity, as required for experiments such as AWAKE, heat-pipe oven concepts are used \cite{Muggli1999}. 
Alkali metal is heated until it vaporises in a pipe.
To confine the vapour cooling jackets are placed at each end of the pipe which cause the vapour to condense, where it is removed by the wick.
Buffer volumes of helium, which has a much higher ionisation energy, further confine the alkali vapour.
Ionisation is done by a laser pulse focused with a long focal length optic or an axicon lens, or by the electric field of a drive beam. 
The latter effect is simpler to utilise in Alkali vapour ovens owing to the intrinsically low ionisation threshold of Alkali metals. 
In a recent variation the vapour is contained by fast valves instead of a buffer gas \cite{Oz2014}.

\subsection{Plasma sources as waveguides}
\label{sec:plasmawaveguides}
LWFA applications usually involve ultrashort laser pulses with peak intensities on the $10^{18}-10^{20}$\,\si{\W\per\square\centi\metre} scale. 
As discussed in Section \ref{para:laserwaveguides} strategies are required to guide the laser pulse at a small spot size, ensuring high intensities, over many Rayleigh lengths.
While this can be done via self-guiding, the highest energy gains from LWFAs to date have employed radial plasma structures to guide the pulse \cite{Leemans2014,Gonsalves2019}.

The change in refractive index from vacuum/gas/plasma to a boundary wall under small grazing incidence angles can guide laser pulses over many Rayleigh lengths. Hollow dielectric capillary tubes \cite{Dorchies1999} have been use to demonstrate monomode guiding of over 100 Rayleigh lengths for 100 fs laser pulses of $10^{16}$\,\si{\W\per\square\centi\metre} intensities. 

Plasma waveguides are another way to overcome the diffraction restriction \cite{Mori1997, Sprangle1987}. A plasma with a transverse electron density profile with a minimum on the axis of propagation, corresponding to a radially decreasing refractive index with a maximum on axis, providing a focusing effect analogous to a lens (see also Section \ref{para:laserwaveguides}). Two different methods are discussed here for creating such a plasma waveguide: laser-heated hydrodynamic expansion, and capillary discharges.

\subsubsection{Hydrodynamic expansion waveguides}
\label{para:hydroewaveguides}

\begin{figure}[t]
    \centering
    \includegraphics[width=0.95\linewidth]{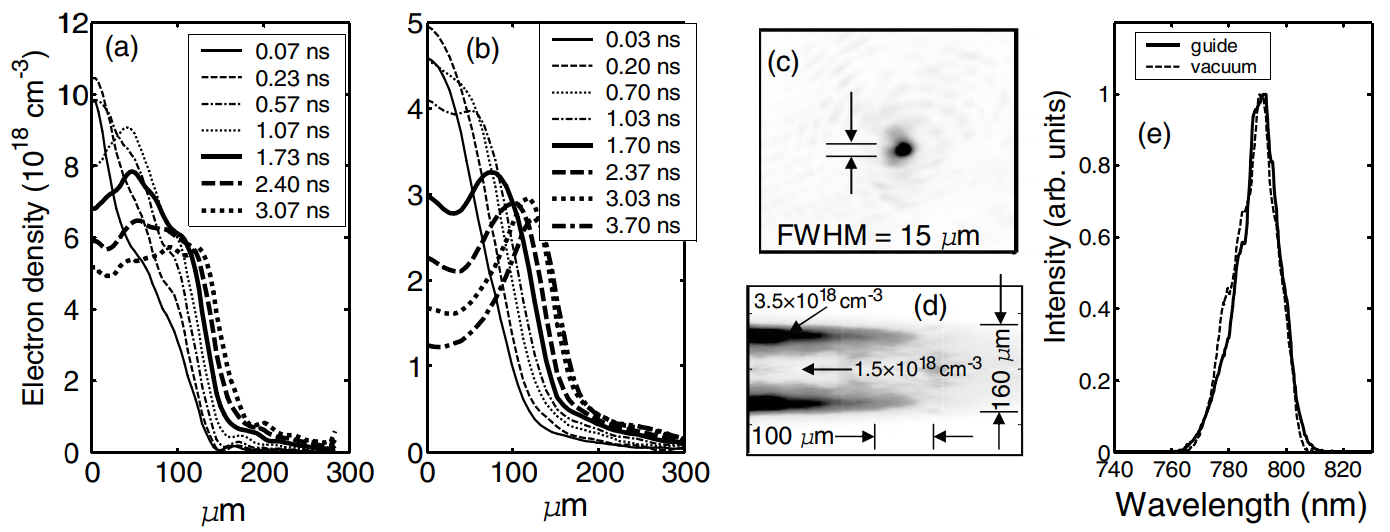}
    \caption{(a), (b) Electron density profiles from 8 mm long plasma channels, with the argon gas jet at 193 K and 27~bar, and 113 K and 20 bar, respectively. (c) Laser mode at the exit of the waveguide (173 K, 40 bar, 1.3 ns delay). (d) Electron density profile at the
waveguide entrance (150 K, 13 bar jet). Note the $\approx$100 $\mu$m length scale over which the waveguide develops from the edge of the gas jet. (e) Guided pulse spectrum at optimum delay (193~K, 27 bar). The spectrum shows negligible additional ionisation by the guided pulse in the waveguide. Image from Ref. \cite{Kumarappan2005} (CC BY 4.0).}
    \label{fig:HydroDens}
\end{figure}

In hydrodynamically expanded plasma channels \cite{Durfee1993,Kumarappan2005} a cylindrical column of plasma is formed and heated by a laser pulse (or a series of laser pulses). The large pressure gradient at the sharp ionised-column boundary drives a radial shock into the surrounding neutral gas, as shown in Fig. \ref{fig:HydroDens}. This rapid expansion of the plasma column leaves behind a depression in the plasma density on-axis behind the~shock. The resulting transverse electron density profile increases radially to the position of the shock front, and hence a refractive index profile is formed which can focus a beam. 
To date the initial plasma column in hydrodynamic
channels has been heated by laser-driven electron-ion
collisions. However, since rapid collisional heating requires high plasma densities it has proved difficult to
generate channels with low on-axis densities. Channels depths of the order of $10^{17}-10^{18}$\,cm$^{-3}$ have been generated in argon and hydrogen plasmas \cite{Kumarappan2005}. Elliptically polarised laser pulses have been demonstrated to control the mean energy of the ionised electrons and hence the development of the channel \cite{Shalloo2019}.

\subsubsection{Discharge waveguides}
\label{para:dischargewaveguides}

A gas-filled capillary discharge waveguide consists of a gas-filled capillary (with pressures of tens to hundreds of millibar) subject to a longitudinal current pulse of a few hundred amperes. This setup is akin to the plasma sources and active plasma lenses. Capillary discharges are an attractive method for forming a plasma waveguide since they require no auxiliary laser, can be scaled to long lengths, and can offer long device lifetimes. In addition, for H-filled capillaries the plasma channel may be fully ionised, which minimises spectral or temporal distortion of the guided laser pulse. 

As discussed in detail in Section \ref{para:temperature}, the balance between resistive heating in the bulk of the~plasma and heat exchange/cooling of the plasma at the capillary walls leads to a non-uniform radial temperature gradient. The associated pressure gradient causes a re-distribution of the plasma density until the pressure is approximately constant along the capillary radius. Under steady-state conditions it can be shown that the electrons density is \cite{Bobrova2001}
\begin{align}
    \frac{n_e(r)}{n_e(0)} \approx \left(1 + 0.33\frac{r^2}{R_0^2}+\dots \right),
\end{align}
and hence near the capillary axis the electron density is approximately parabolic. Here $R_0$ is the capillary radius and $n_e(0)$ is the on-axis electron density. It should be noted that this relation depends only upon the assumption that steady-state conditions have been reached, and does not depend on the current used or temperature conditions reached to achieve the steady-state. The matched spot size is then
\begin{align}
    w_m \approx \left(\frac{R_0^2}{0.33 \pi r_e n_e(0)}\right)^{1/4},
\end{align}
where $r_e$ is the classical electron radius. Capillary discharge waveguides have been successfully employed to guide laser pulses with peak intensities greater than $10^{17}$ W/cm$^2$ over distances greater than 50~mm \cite{Butler2002}.

Laser heating can be used in addition to the discharge \cite{Bobrova2013}. Once the discharge-initiated parabolic density is formed in the capillary, a long (ns) laser pulse is used to additionally heat the electrons on the~capillary axis. The resulting profile then has a second, deeper channel on axis, the properties of which (width and depth) can be controlled by the heating laser. The deeper channel allows laser guiding at low plasma densities. This is of highest importance for high-peak power systems in the Petawatt regime \cite{Gonsalves2019}.

\subsection{Active plasma lenses}
\label{para:activeplasmalens}

\begin{figure}[t]
    \centering
    \includegraphics[width=0.95\linewidth]{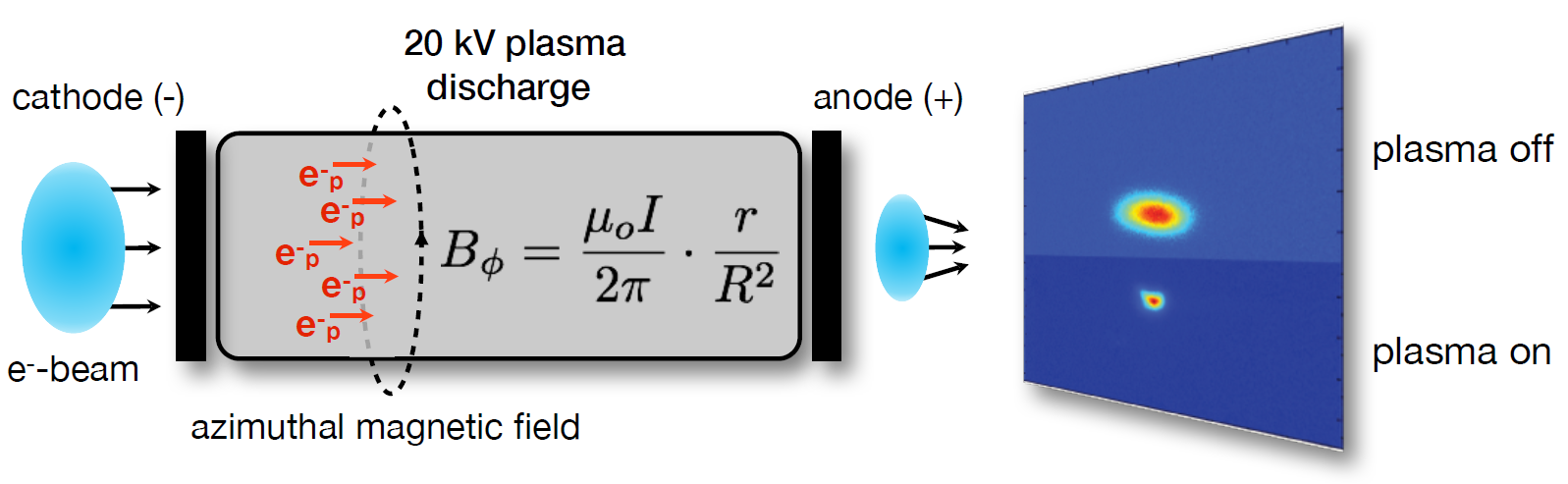}
    \caption{Diagram of an active plasma lens. The current discharge induces an azimuthal magnetic field which can be used to focus an electron beam.}
    \label{fig:APL}
\end{figure}

In addition to creating the plasma, the discharge itself can be exploited to create specialised charged-particle focusing optics known as active plasma lenses (APLs). Active plasma lenses are a promising method for strong focusing of particle beams as they provide tuneable radially-symmetric kT/m focusing fields \cite{vanTilborg2017,Lindstrm2018}. An active plasma lens consists of a thin gas-filled capillary with high voltage electrodes either side supplying the discharge current (see Fig. \ref{fig:APL}).
When the gas is discharged, a longitudinal current flows through the capillary. Ampere's law states that
\begin{align}
    \nabla \times B = \mu_0 J,
\end{align}
where $B$ is the magnetic (vector) field, $J$ is the current density and $\mu_0$ is the permeability of free space. Assuming cylindrical symmetry and only a longitudinal current, Ampere's law can be simplified to
\begin{align}
    \frac{1}{r}\frac{\partial}{\partial r}\left( rB_\phi \right) = \mu_0J_z(r), \label{eq:AmpCyl}
\end{align}
where r, $\phi$ and z are the radial, azimuthal and longitudinal coordinates. 

The advantages of APLs should now be clear. The focusing can be achieved in both horizontal and vertical planes simultaneously without the need for much larger systems of quadrupoles/solenoids. The~magnetic focusing strength is directly controllable by changing the applied current. Finally, focusing gradients of $\approx 5$ kT/m have been recently demonstrated, which is an order of magnitude greater than that produced by magnetic quadrupoles \cite{Weingartner2011}. APLs played a critical role in demonstrating the staging of two LWFA components \cite{Steinke2016}.

\begin{figure}[t]
    \centering
    \includegraphics[width=0.98\columnwidth]{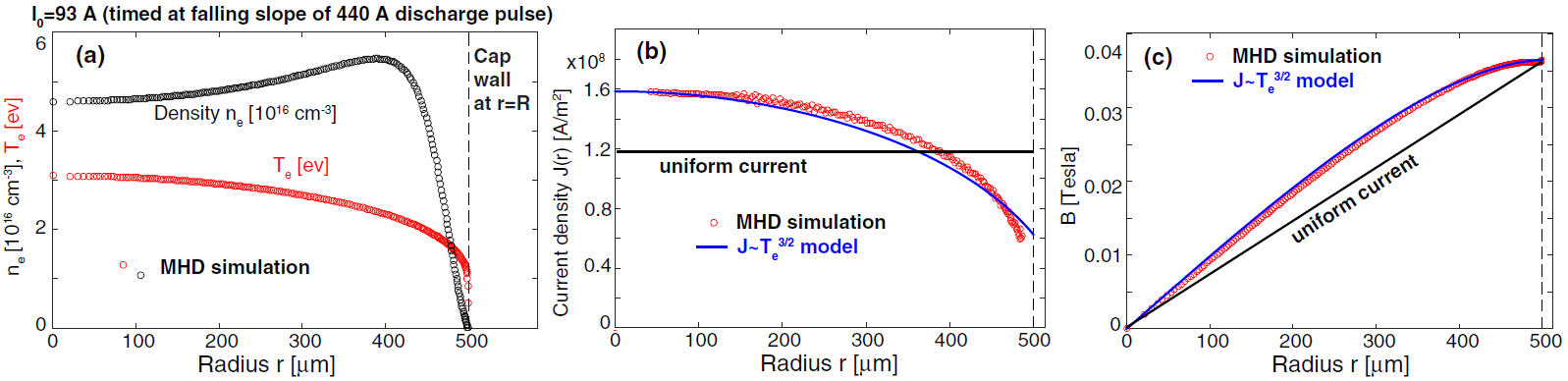}
    \caption{MHD simulations of a 1-mm-diameter plasma lens
yield the radial distribution of (a) the electron temperature and plasma density, (b) the transverse current density, and (c) the magnetic field profile. A simplified model was proposed (plotted as solid blue curves), based on $J(r)\propto T_e^{3/2}$ as described in \cite{Bobrova2001}. Image from Ref.~\cite{vanTilborg2017} (CC BY 4.0).}
    \label{fig:Tilborg}
\end{figure}

\begin{figure}[h!]
    \centering
    \includegraphics[width=0.9\linewidth]{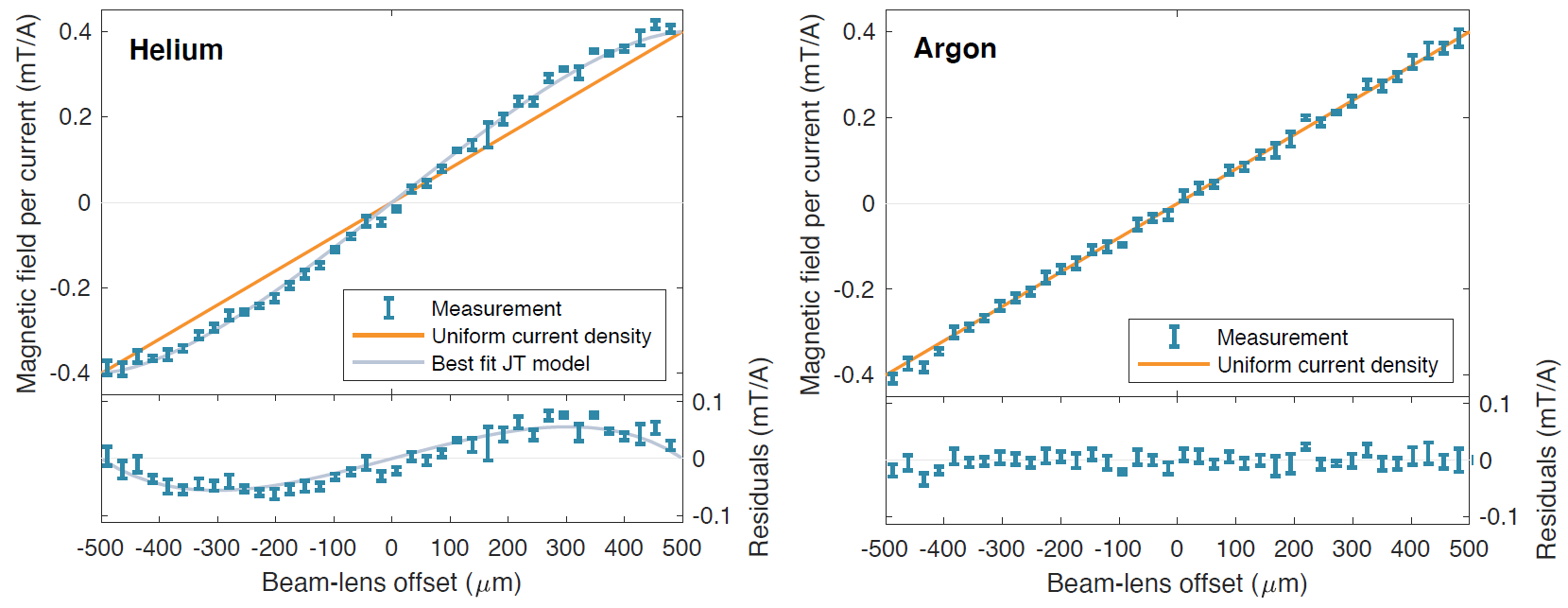}
    \caption{Measurement of the magnetic field per discharge current for a scan of beam-to-lens offsets in helium and argon. A strong nonlinearity is observed in helium, whereas in argon the measurement is consistent with the~expectation from a uniform current density. Image from Ref. \cite{Lindstrm2018} (CC BY 4.0).}
    \label{fig:Lindstrom}
\end{figure}

Emittance is preserved (with respect to the electromagnetic focusing, i.e., ignoring chromatic effects or background collisional effects) for a linear radial magnetic field  \cite{vanTilborg2018}. For a uniform current density, $J_{ideal} = I/(\pi R^2)$ where $I$ is the total current, Eq. (\ref{eq:AmpCyl}) can be integrated to give
\begin{align}
    B_{ideal} &= \frac{\mu_0}{2}rJ_{ideal} = \frac{\mu_0 I}{2\pi R^2}r.
\end{align}
Thus an ideal plasma lens, i.e., one with an emittance-preserving magnetic field, is achieved for uniform radial current density. From Ohm's law we have,
\begin{align}
    J_z(r) = \sigma(r)E_z
\end{align}
where $\sigma$ is the electrical conductivity, i.e. the inverse of the electrical resistivity $\eta = \frac{1}{\sigma}$. For an ideal plasma, the electrical conductivity is given by the Spitzer relation \cite{atzeni2004}
\begin{align}
    \sigma &\propto \frac{T_e^{3/2}}{\ln{\Lambda_e\left(n_e,T_e\right)}}
\end{align}
which depends strongly on electron temperature $T_e$ and weakly on electron density $n_e$ via the coulomb logarithm $\ln{\Lambda_e}$. Clearly, a uniform current density requires a uniform radial plasma density and temperature, but as already discussed, the balance between resistive heating and wall-cooling can give rise to significant temperature and density gradients within the plasma. The effect of the non-linear plasma properties on the magnetic field is shown for both simulation (Fig.~\ref{fig:Tilborg}) and real experimental data (Fig.~\ref{fig:Lindstrom}).

The larger the deviation of the magnetic field profile from the ideal linear case, the~larger the~emittance degradation. 
In practice, one ideally will operate an APL at a time before the significant temperature and density gradients develop, which varies significantly between different ion species. For example, light gases such as hydrogen and helium respond much faster to changes in conditions than heavier gases such as argon. A comparison of the magnetic field profile $\approx 80$ ns after the initiation of a~current discharge between helium and argon plasma capillaries is shown in Fig.~\ref{fig:Lindstrom}.
In this example, the~emittance is not conserved to within experimental error for helium, whereas it is for argon.

\section{Plasma diagnostics}

\begin{figure}
\centering
	\includegraphics[width=0.45\linewidth]{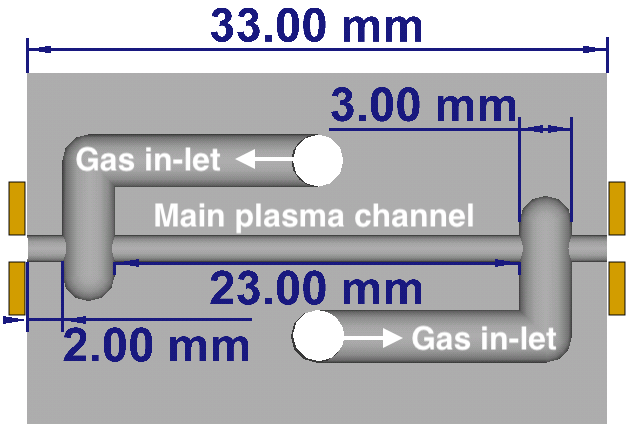}
	
	\caption{A schematic of the type of capillary discharge used at FLASHForward. The central channel and gas inlets have a circular cross section and are milled from sapphire. The exits of the main channel are open to allow continuous gas flow.}
	\label{fig:CapExp}
\end{figure}

Capillary discharge plasma sources~\cite{gonsalves2011tunable,bendoyro2008plasma} are widely used for laser-driven and beam-driven plasma accelerator applications. Figure~\ref{fig:CapExp} displays a schematic of the type of capillary discharge used at the~PWFA facility FLASHForward~\cite{d2019flashforward}. This type of source has a cylindrical channel milled from sapphire with gas inlets through which a continuous flow of gas is fed. At the capillary extremities are electrodes for enabling the discharge current, which are cut with a hole of the same bore diameter as the capillary cross section. The sapphire material allows the transmission of light emitted during the creation and recombination of the plasma and hence enables diagnostic spectroscopy to be performed. Additionally the~open-ended capillary geometry facilitates the passage of a laser diagnostic along the longitudinal axis.

\subsection{Gas density distribution}

The initial neutral gas density and its spatial distribution within a gas-filled plasma source governs the~resulting plasma density. It is therefore interesting to measure the neutral gas density in order to better understand the initial plasma density distribution. This is commonly realised by transverse laser interferometry (cf.~to Section \ref{laserint} for details). A more sensitive, but not widely used method is Raman spectroscopy of molecular gases. This process utilises inelastic scattering of the laser light from vibrational and rotational molecular states and can distinguish between gas species~\cite{schaper2014longitudinal}. 

\subsection{Plasma density distribution}

Knowledge of the plasma density distribution and its spatial and temporal evolution are of great importance in plasma accelerators in order to be able to realise controlled wakefield creation and acceleration. Therefore, sensitive diagnostics capable of resolving spatial and temporal plasma density have been developed in various capacities. Below we describe two types of techniques relevant to plasma density diagnosis in discharge capillary sources.

\subsubsection{Plasma spectroscopy}

\begin{figure}[t]
\centering
\includegraphics[width=0.48\columnwidth]{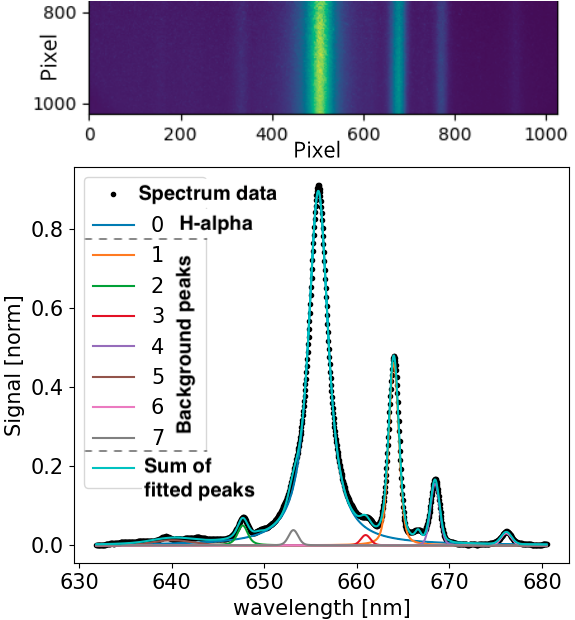}
\caption{The raw spectrometer image from plasma light in argon doped with hydrogen is shown in the upper part of the figure. The lower part shows the resulting projected spectrum (black points) with the dominant $H_{\alpha}$ emission line in the centre. Each individual background peak is identified and fitted as well as a sum of all the fitted peaks in light blue.}
\label{fig:ArSpec}
\end{figure}

The spectrum of light emitted by a plasma can be used to obtain information about the plasma density. One of the most widely used techniques is the analysis of spectral line broadening due to the Stark effect; the splitting of atomic decay states due to the local electric field strength, from in this case a~plasma. By utilising methodology developed by for example Gigosos and Cardenoso (GC)~\cite{gigosos1996new,gigosos2003computer}, the plasma density may be calculated by measuring the full width at half maximum (FWHM) $\Delta\lambda$ of a given broadened spectral line emitted by atoms in a plasma. Figure~\ref{fig:ArSpec} shows an example of the broadening of the $H_{\alpha}$ emission line in the Balmer series for the case where argon gas is doped with hydrogen. Care must be taken to fit all background peaks in such a spectrum as shown in Fig.~\ref{fig:ArSpec} and hence extract the information about the desired peak. Following the technique developed by GC, the plasma density can then be calculated from the value $\Delta\lambda$ of such a profile, as
\begin{equation}
	n_e = A\cdot \Delta\lambda^{B},
\label{eq:specDens}
\end{equation}
where $A$ and $B$ are constants derived from a fit to simulation data benchmarked against experimental measurements.

Without precise knowledge of the plasma constituent temperatures, an uncertainty of 20~-~25\% exists on the measurement of the plasma density. However, Stark broadening measurements can be benchmarked against other techniques to reduce the uncertainty. In this type of measurement the spectral, spatial and temporal resolution depend mostly on the properties of the spectrometer and camera equipment used. For example a fast-gated camera with an intensified CCD chip (iCCD) can increase the temporal resolution greatly by allowing light to be collected in the nanosecond regime. A balance between light collection and resolution must be established in such cases. 

\subsubsection{Laser interferometry} \label{laserint}

\begin{figure}[t]
    \centering
	\includegraphics[width=0.6\linewidth]{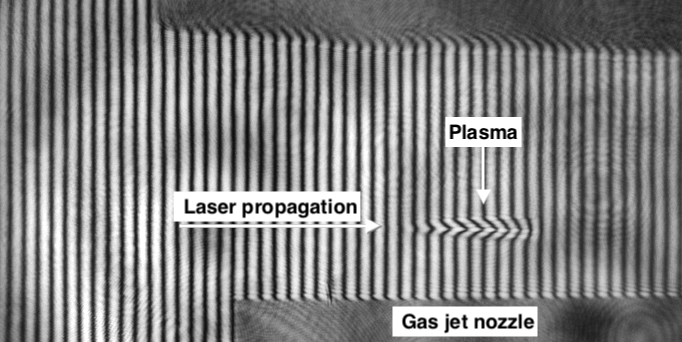}
	\caption{The interference pattern created using a Michelson interferometer to observe a plasma created by a laser and gas jet.}
	\label{fig:TMLI}
\end{figure}

Laser interferometry provides the possibility for fast and accurate line-of-sight integrated plasma density measurements without, for example, uncertainties due to temperature. The integrated path length or phase of a laser pulse traversing a plasma depends on the index of refraction of the plasma,
\begin{equation}
	\eta=\sqrt{1-\left(\frac{\omega_p}{\omega_0}\right)^2},
\label{eq:Plasrefrac}
\end{equation}
where $\omega_p$ and $\omega_0$ are the plasma frequency and the fundamental frequency of the laser pulse respectively. The interference created by combining different laser arms of a spectrometer where one arm has traversed a plasma and one has not, allows for the direct calculation of the average plasma density. Figure~\ref{fig:TMLI} shows the interference pattern created using a Mach-Zehnder interferometer to observe fringes created from a~laser-ionised plasma. This interference pattern must be referenced against one without a plasma and then the total phase shift $\Delta\phi$ may be calculated. For a given laser frequency $\omega_0$, Eq.~(\ref{eq:Plasrefrac}) can be rearranged in terms of the plasma density to yield
\begin{equation}
	n_e = \frac{4 c \epsilon_0 m_e}{ q_e^2}\frac{\omega_0}{L}\Delta\phi,
\label{eq:TALI}
\end{equation}
where $\omega_0$ is the fundamental wavelength of the laser pulse, $L$ is the length of plasma traversed and $\Delta\phi$ is the total integrated phase shift. The fundamental constants $c$, $\epsilon_0$, $m_e$ and $q_e$ have their usual meanings. 

Such a measurement technique suffers from the limitation that the total phase shift observed is a~line-of-sight integral and therefore the geometry of the target must be considered. Due to the fact that most plasma targets have cylindrical symmetry, the technique of Abel inversion can often be used to resolve the spatial density profile~\cite{kalal1988abel}.

\begin{figure}[t]
    \centering
	\includegraphics[width=0.95\linewidth]{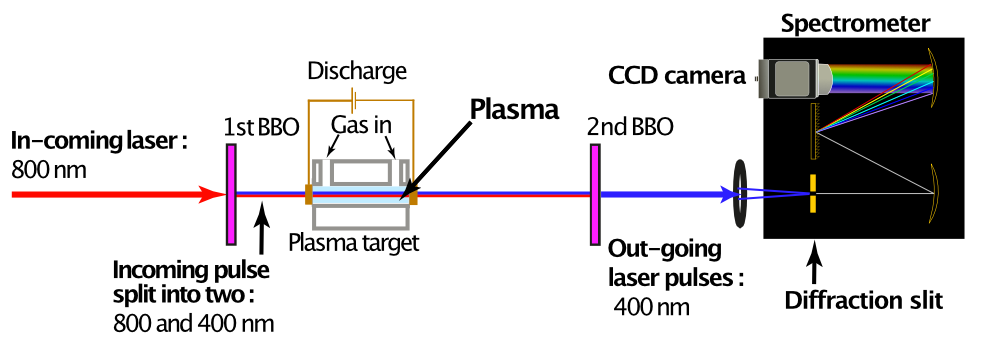}
	\caption{Schematic of a longitudinal two-colour laser interferometer constructed at DESY, Hamburg.}
	\label{fig:TCI}
\end{figure}

\begin{figure}[t]
\centering
\subfigure[]{
	\includegraphics[width=0.5\columnwidth]{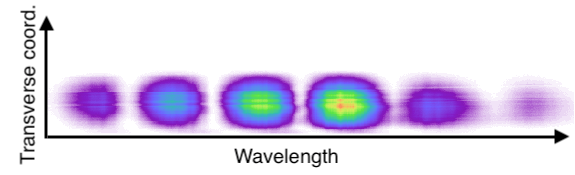}
	}
\subfigure[]{
	\includegraphics[width=0.5\columnwidth]{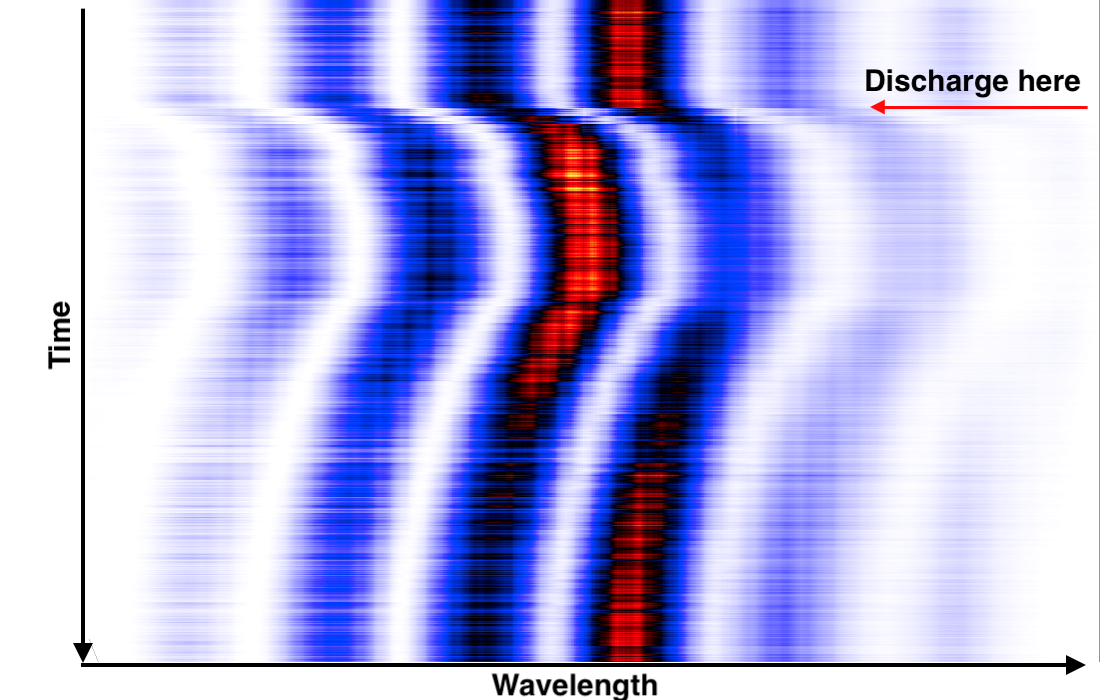}
	}	
\caption{(a) A typical interference pattern observed on the camera after the slit and spectrometer. (b) Waterfall plot of a typical interference pattern projected onto the wavelength axis as a function of time.} 
\label{fig:TCI_intPat+waterfall}
\end{figure}

A recent technique~\cite{van2019density} employs a common path two-colour laser interferometer (TCI) to measure the group and/or phase velocity delay accrued by two harmonic pulses of the same laser in a plasma as shown in Fig.~\ref{fig:TCI}.
A typical interference pattern is shown in Fig.~\ref{fig:TCI_intPat+waterfall} along with a waterfall plot showing the temporal change in the interference pattern created by the phase accumulation difference of two such harmonics of a single laser pulse during plasma recombination. The temporal plasma density evolution can be calculated by measuring the accumulated phase shift shown in Fig.~\ref{fig:TCI_intPat+waterfall}~(b) by modifying Eq.~(\ref{eq:TALI}) to take account for the first and second harmonic of the laser frequency which introduces an additional factor of 1/3 to the right hand side.

\begin{figure}[t]
    \centering
	\includegraphics[width=0.65\linewidth]{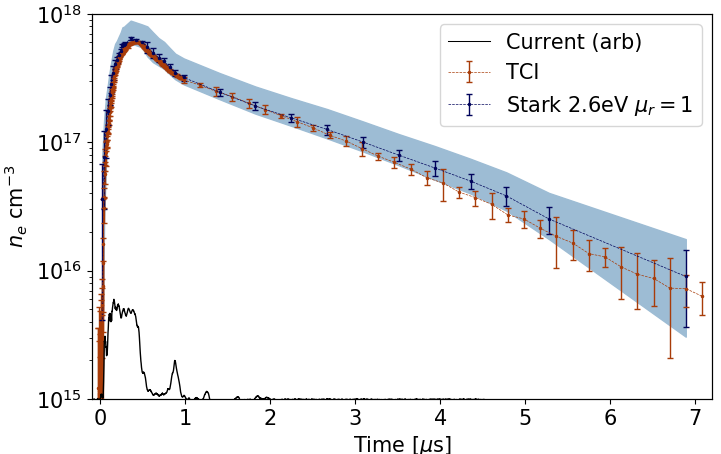}
	\caption{Electrical discharge-ignited plasma density measurements recorded using a TCI (orange) and Stark broadened hydrogen emission profiles (blue). The blue band shows the uncertainty due to the unknown plasma temperature in the Stark-broadening process. The TCI measurement is temperature independent.}
	\label{fig:TCI_Spec_comp}
\end{figure}

In a plasma source such as that shown in Fig.~\ref{fig:CapExp} it has been demonstrated~\cite{garland2020characterisation} that when using optical emission spectroscopy and two-colour laser interferometry it is possible to acquire accurate temporally and spatially resolved plasma density information as shown in Fig.~\ref{fig:TCI_Spec_comp}.


\pagebreak
\section{Summary}
Tailoring, controlling, and diagnosing the properties of plasma sources is paramount to accurately design, manipulate, and operate plasma wakefield accelerators. Plasma density and temperature profiles in all dimensions dictate or, at least, influence every aspect of the plasma acceleration process. In this article, we have reviewed the underlying physics concepts and introduced some of the most fundamental schemes to provide the basis for advanced studies of the highly dynamic field of plasma-based particle acceleration. Plasma generation is generally achieved by a combination of electrical discharges, laser ionisation and beam ionisation, and by careful consideration of the resulting plasma density, shape, temperature etc., estimates of the optimum conditions for plasma wakefield acceleration (and associated applications, such as active plasma lenses) can be made. An appropriate plasma acceleration source will include input and output matching, trapping, and acceleration stages, all which require individual tailoring. The technical implementation of plasma sources via gas jets, gas cells, heat pipe ovens, and the development of plasma waveguides have all been considered. Finally, diagnostic techniques including plasma spectroscopy and laser interferometry have been discussed and compared for measuring the electron density on plasma experiments, which provides crucial feedback for the tailoring and optimisation of plasma acceleration sources.

\bibliography{refs}

\begin{thebibliography}{10}

\bibitem{Fair2014}
D.C. Faircloth.
\newblock Technological aspects: High voltage, April 2014.

\bibitem{KunLue2013}
Erich~E. Kunhardt and Lawrence~H. Luessen.
\newblock {\em Electrical Breakdown and Discharges in Gases - Part A
  Fundamental Processes and Breakdown}.
\newblock Springer Science and Business Media, Berlin Heidelberg, 2013.

\bibitem{LiebLich2005}
Michael~A. Lieberman.
\newblock {\em Principles of Plasma Discharges and Materials Processing (2nd
  ed.)}.
\newblock Wiley-Interscience, 2005.

\bibitem{Gibbon}
P.~Gibbon.
\newblock {\em Short Pulse Laser Interactions with Matter: An Introduction}.
\newblock Imperial College Press, 2005.

\bibitem{ADK}
M.~V. Ammosov, N.~B. Delone, and V.~P. Krainov.
\newblock Tunnel ionization of complex atoms and of atomic ions in an
  alternating electromagnetic field.
\newblock {\em {JETP}}, 64(6), December 1986.

\bibitem{Tong_2005}
X~M Tong and C~D Lin.
\newblock Empirical formula for static field ionization rates of atoms and
  molecules by lasers in the barrier-suppression regime.
\newblock {\em Journal of Physics B: Atomic, Molecular and Optical Physics},
  38(15):2593--2600, July 2005.

\bibitem{Zhang2014}
Qingbin Zhang, Pengfei Lan, and Peixiang Lu.
\newblock Empirical formula for over-barrier strong-field ionization.
\newblock {\em Phys. Rev. A}, 90:043410, October 2014.

\bibitem{Lu2005}
W.~Lu, C.~Huang, M.~M. Zhou, W.~B. Mori, and T.~Katsouleas.
\newblock Limits of linear plasma wakefield theory for electron or positron
  beams.
\newblock {\em Physics of Plasmas}, 12(6):063101, 2005.

\bibitem{Lu2010}
W.~Lu, W.~An, M.~Zhou, C.~Joshi, C.~Huang, and W.~B. Mori.
\newblock The optimum plasma density for plasma wakefield excitation in the
  blowout regime w.
\newblock {\em New Journal of Physics}, 12:085002, 2010.

\bibitem{Lu2007}
W.~Lu, M.~Tzoufras, C.~Joshi, F.~S. Tsung, W.~B. Mori, J.~Vieira, R.~A.
  Fonseca, and L.~O. Silva.
\newblock Generating multi-gev electron bunches using single stage laser
  wakefield acceleration in a 3d nonlinear regime.
\newblock {\em Phys. Rev. ST Accel. Beams}, 10:061301, June 2007.

\bibitem{Litos2019}
M.~D. Litos, R.~Ariniello, C.~E. Doss, K.~Hunt-Stone, and J.~R. Cary.
\newblock Beam emittance preservation using gaussian density ramps in a
  beam-driven plasma wakefield accelerator.
\newblock {\em Philosophical Transactions of the Royal Society A: Mathematical,
  Physical and Engineering Sciences}, 377(2151):20180181, 2019.

\bibitem{Bourgeois2013}
N.~Bourgeois, J.~Cowley, and S.~M. Hooker.
\newblock Two-pulse ionization injection into quasilinear laser wakefields.
\newblock {\em Phys. Rev. Lett.}, 111:155004, October 2013.

\bibitem{poderthesis}
Kristjan Poder.
\newblock {\em {Characterisation of self-guided laser wakefield accelerators to
  multi-GeV energies}}.
\newblock PhD thesis, Imperial College London, 2017.

\bibitem{Pollock2011}
B.~B. Pollock, C.~E. Clayton, J.~E. Ralph, F.~Albert, A.~Davidson, L.~Divol,
  C.~Filip, S.~H. Glenzer, K.~Herpoldt, W.~Lu, K.~A. Marsh, J.~Meinecke, W.~B.
  Mori, A.~Pak, T.~C. Rensink, J.~S. Ross, J.~Shaw, G.~R. Tynan, C.~Joshi, and
  D.~H. Froula.
\newblock Demonstration of a narrow energy spread, $\ensuremath{\sim}0.5\text{
  }\text{ }\mathrm{GeV}$ electron beam from a two-stage laser wakefield
  accelerator.
\newblock {\em Phys. Rev. Lett.}, 107:045001, July 2011.

\bibitem{Kim2013}
Hyung~Taek Kim, Ki~Hong Pae, Hyuk~Jin Cha, I~Jong Kim, Tae~Jun Yu, Jae~Hee
  Sung, Seong~Ku Lee, Tae~Moon Jeong, and Jongmin Lee.
\newblock Enhancement of electron energy to the multi-gev regime by a
  dual-stage laser-wakefield accelerator pumped by petawatt laser pulses.
\newblock {\em Phys. Rev. Lett.}, 111:165002, October 2013.

\bibitem{Rosenzweig2004}
J.~B. Rosenzweig, N.~Barov, M.~C. Thompson, and R.~B. Yoder.
\newblock Energy loss of a high charge bunched electron beam in plasma:
  Simulations, scaling, and accelerating wakefields.
\newblock {\em Phys. Rev. ST Accel. Beams}, 7:061302, June 2004.

\bibitem{Lotov2007}
K.~V. Lotov and V.~S. Tikhanovich.
\newblock Numerical optimization of a plasma wakefield acceleration experiment.
\newblock {\em Phys. Rev. ST Accel. Beams}, 10:051301, May 2007.

\bibitem{Mori1997}
W.~B. {Mori}.
\newblock The physics of the nonlinear optics of plasmas at relativistic
  intensities for short-pulse lasers.
\newblock {\em IEEE Journal of Quantum Electronics}, 33(11):1942--1953,
  November 1997.

\bibitem{Sprangle1987}
P.~{Sprangle}, C.~{Tang}, and E.~{Esarey}.
\newblock Relativistic self-focusing of short-pulse radiation beams in plasmas.
\newblock {\em IEEE Transactions on Plasma Science}, 15(2):145--153, April
  1987.

\bibitem{Poder_2017}
K~Poder, J~M Cole, J~C Wood, N~C Lopes, S~Alatabi, P~S Foster, C~Kamperidis,
  O~Kononenko, C~A Palmer, D~Rusby, A~Sahai, G~Sarri, D~R Symes, J~R Warwick,
  S~P~D Mangles, and Z~Najmudin.
\newblock Measurements of self-guiding of ultrashort laser pulses over long
  distances.
\newblock {\em Plasma Physics and Controlled Fusion}, 60(1):014022, October
  2017.

\bibitem{Kneip2009}
S.~Kneip, S.~R. Nagel, S.~F. Martins, S.~P.~D. Mangles, C.~Bellei, O.~Chekhlov,
  R.~J. Clarke, N.~Delerue, E.~J. Divall, G.~Doucas, K.~Ertel, F.~Fiuza,
  R.~Fonseca, P.~Foster, S.~J. Hawkes, C.~J. Hooker, K.~Krushelnick, W.~B.
  Mori, C.~A.~J. Palmer, K.~Ta Phuoc, P.~P. Rajeev, J.~Schreiber, M.~J.~V.
  Streeter, D.~Urner, J.~Vieira, L.~O. Silva, and Z.~Najmudin.
\newblock Near-gev acceleration of electrons by a nonlinear plasma wave driven
  by a self-guided laser pulse.
\newblock {\em Phys. Rev. Lett.}, 103:035002, July 2009.

\bibitem{Clayton2010}
C.~E. Clayton, J.~E. Ralph, F.~Albert, R.~A. Fonseca, S.~H. Glenzer, C.~Joshi,
  W.~Lu, K.~A. Marsh, S.~F. Martins, W.~B. Mori, A.~Pak, F.~S. Tsung, B.~B.
  Pollock, J.~S. Ross, L.~O. Silva, and D.~H. Froula.
\newblock Self-guided laser wakefield acceleration beyond 1 gev using
  ionization-induced injection.
\newblock {\em Phys. Rev. Lett.}, 105:105003, September 2010.

\bibitem{Wang2013}
X.~Wang, R.~Zgadzaj, N.~Fazel, Z.~Li, S.~A. Yi, Xi~Zhang, W.~Henderson, Y.-Y.
  Chang, R.~Korzekwa, H.-E. Tsai, C.-H. Pai, H.~Quevdeo, G.~Dyer, E.~Gaul,
  M.~Martinez, A.~C. Bernstein, T.~Borger, M.~Spinks, M.~Donovan, V.~Khudik,
  G.~Shvets, T.~Ditmire, and M.~C. Downer.
\newblock Quasi-monoenergetic laser-plasma acceleration of electrons to 2\,gev.
\newblock {\em Nature Communications}, 4(1988), 2013.

\bibitem{Leemans2014}
W.~P. Leemans, A.~J. Gonsalves, H.-S. Mao, K.~Nakamura, C.~Benedetti, C.~B.
  Schroeder, Cs. T\'oth, J.~Daniels, D.~E. Mittelberger, S.~S. Bulanov, J.-L.
  Vay, C.~G.~R. Geddes, and E.~Esarey.
\newblock Multi-gev electron beams from capillary-discharge-guided subpetawatt
  laser pulses in the self-trapping regime.
\newblock {\em Phys. Rev. Lett.}, 113:245002, December 2014.

\bibitem{Gonsalves2019}
A.~J. Gonsalves, K.~Nakamura, J.~Daniels, C.~Benedetti, C.~Pieronek, T.~C.~H.
  de~Raadt, S.~Steinke, J.~H. Bin, S.~S. Bulanov, J.~van Tilborg, C.~G.~R.
  Geddes, C.~B. Schroeder, Cs. T\'oth, E.~Esarey, K.~Swanson, L.~Fan-Chiang,
  G.~Bagdasarov, N.~Bobrova, V.~Gasilov, G.~Korn, P.~Sasorov, and W.~P.
  Leemans.
\newblock Petawatt laser guiding and electron beam acceleration to 8 gev in a
  laser-heated capillary discharge waveguide.
\newblock {\em Phys. Rev. Lett.}, 122:084801, February 2019.

\bibitem{Esarey1997}
E.~Esarey, P.~Sprangle, J.~Krall, and A.~Ting.
\newblock Self-focusing and guiding of short laser pulses in ionizing gases and
  plasmas.
\newblock {\em IEEE Journal of Quantum Electronics}, 33:1879--1914, Nov 1997.

\bibitem{Spence2000}
D.~J. Spence and S.~M. Hooker.
\newblock Investigation of a hydrogen plasma waveguide.
\newblock {\em Phys. Rev. E}, 63:015401, December 2000.

\bibitem{Butler2002}
A.~Butler, D.~J. Spence, and S.~M. Hooker.
\newblock Guiding of high-intensity laser pulses with a hydrogen-filled
  capillary discharge waveguide.
\newblock {\em Phys. Rev. Lett.}, 89:185003, October 2002.

\bibitem{Durfee1993}
C.~G. Durfee and H.~M. Milchberg.
\newblock Light pipe for high intensity laser pulses.
\newblock {\em Phys. Rev. Lett.}, 71:2409--2412, October 1993.

\bibitem{Lemos2013}
N.~Lemos, T.~Grismayer, L.~Cardoso, G.~Figueira, R.~Issac, D.~A. Jaroszynski,
  and J.~M. Dias.
\newblock Plasma expansion into a waveguide created by a linearly polarized
  femtosecond laser pulse.
\newblock {\em Physics of Plasmas}, 20(6):063102, 2013.

\bibitem{Shalloo2019}
R.~J. Shalloo, C.~Arran, A.~Picksley, A.~von Boetticher, L.~Corner,
  J.~Holloway, G.~Hine, J.~Jonnerby, H.~M. Milchberg, C.~Thornton, R.~Walczak,
  and S.~M. Hooker.
\newblock Low-density hydrodynamic optical-field-ionized plasma channels
  generated with an axicon lens.
\newblock {\em Phys. Rev. Accel. Beams}, 22:041302, April 2019.

\bibitem{Esarey1995}
Eric Esarey and Mark Pilloff.
\newblock Trapping and acceleration in nonlinear plasma waves.
\newblock {\em Physics of Plasmas}, 2(5):1432--1436, 1995.

\bibitem{Bulanov1998}
S.~Bulanov, N.~Naumova, F.~Pegoraro, and J.~Sakai.
\newblock Particle injection into the wave acceleration phase due to nonlinear
  wake wave breaking.
\newblock {\em Phys. Rev. E}, 58:R5257--R5260, November 1998.

\bibitem{Geddes2008}
C.~G.~R. Geddes, K.~Nakamura, G.~R. Plateau, Cs. Toth, E.~Cormier-Michel,
  E.~Esarey, C.~B. Schroeder, J.~R. Cary, and W.~P. Leemans.
\newblock Plasma-density-gradient injection of low absolute-momentum-spread
  electron bunches.
\newblock {\em Phys. Rev. Lett.}, 100:215004, May 2008.

\bibitem{DeLaOssa2017}
A.~Martinez de~la Ossa, Z.~Hu, M.~J.~V. Streeter, T.~J. Mehrling, O.~Kononenko,
  B.~Sheeran, and J.~Osterhoff.
\newblock Optimizing density down-ramp injection for beam-driven plasma
  wakefield accelerators.
\newblock {\em Phys. Rev. Accel. Beams}, 20:091301, September 2017.

\bibitem{KONONENKO2016125}
O.~Kononenko, N.C. Lopes, J.M. Cole, C.~Kamperidis, S.P.D. Mangles,
  Z.~Najmudin, J.~Osterhoff, K.~Poder, D.~Rusby, D.R. Symes, J.~Warwick, J.C.
  Wood, and C.A.J. Palmer.
\newblock 2d hydrodynamic simulations of a variable length gas target for
  density down-ramp injection of electrons into a laser wakefield accelerator.
\newblock {\em Nuclear Instruments and Methods in Physics Research Section A:
  Accelerators, Spectrometers, Detectors and Associated Equipment}, 829:125 --
  129, 2016.
\newblock 2nd European Advanced Accelerator Concepts Workshop - EAAC 2015.

\bibitem{WITTIG201683}
Georg Wittig, Oliver~S. Karger, Alexander Knetsch, Yunfeng Xi, Aihua Deng,
  James~B. Rosenzweig, David~L. Bruhwiler, Jonathan Smith, Zheng-Ming Sheng,
  Dino~A. Jaroszynski, Grace~G. Manahan, and Bernhard Hidding.
\newblock Electron beam manipulation, injection and acceleration in plasma
  wakefield accelerators by optically generated plasma density spikes.
\newblock {\em Nuclear Instruments and Methods in Physics Research Section A:
  Accelerators, Spectrometers, Detectors and Associated Equipment}, 829:83 --
  87, 2016.
\newblock 2nd European Advanced Accelerator Concepts Workshop - EAAC 2015.

\bibitem{Suk2001}
H.~Suk, N.~Barov, J.~B. Rosenzweig, and E.~Esarey.
\newblock Plasma electron trapping and acceleration in a plasma wake field
  using a density transition.
\newblock {\em Phys. Rev. Lett.}, 86:1011--1014, February 2001.

\bibitem{Buck2013}
A.~Buck, J.~Wenz, J.~Xu, K.~Khrennikov, K.~Schmid, M.~Heigoldt, J.~M.
  Mikhailova, M.~Geissler, B.~Shen, F.~Krausz, S.~Karsch, and L.~Veisz.
\newblock Shock-front injector for high-quality laser-plasma acceleration.
\newblock {\em Phys. Rev. Lett.}, 110:185006, May 2013.

\bibitem{Rittershofer2010}
W.~Rittershofer, C.~B. Schroeder, E.~Esarey, F.~J. Grüner, and W.~P. Leemans.
\newblock Tapered plasma channels to phase-lock accelerating and focusing
  forces in laser-plasma accelerators.
\newblock {\em Physics of Plasmas}, 17(6):063104, 2010.

\bibitem{Dopp2016}
A.~Döpp, E.~Guillaume, C.~Thaury, A.~Lifschitz, K.~Ta~Phuoc, and V.~Malka.
\newblock Energy boost in laser wakefield accelerators using sharp density
  transitions.
\newblock {\em Physics of Plasmas}, 23(5):056702, 2016.

\bibitem{Floettmann2003}
Klaus Floettmann.
\newblock Some basic features of the beam emittance.
\newblock {\em Phys. Rev. ST Accel. Beams}, 6:034202, March 2003.

\bibitem{MEHRLING2016367}
T.J. Mehrling, R.E. Robson, J-H. Erbe, and J.~Osterhoff.
\newblock Efficient numerical modelling of the emittance evolution of beams
  with finite energy spread in plasma wakefield accelerators.
\newblock {\em Nuclear Instruments and Methods in Physics Research Section A:
  Accelerators, Spectrometers, Detectors and Associated Equipment}, 829:367 --
  371, 2016.
\newblock 2nd European Advanced Accelerator Concepts Workshop - EAAC 2015.

\bibitem{Mehrling2017}
T.~J. Mehrling, R.~A. Fonseca, A.~Martinez de~la Ossa, and J.~Vieira.
\newblock Mitigation of the hose instability in plasma-wakefield accelerators.
\newblock {\em Phys. Rev. Lett.}, 118:174801, April 2017.

\bibitem{Brinkmann2017}
R.~Brinkmann, N.~Delbos, I.~Dornmair, M.~Kirchen, R.~Assmann, C.~Behrens,
  K.~Floettmann, J.~Grebenyuk, M.~Gross, S.~Jalas, T.~Mehrling, A.~Martinez
  de~la Ossa, J.~Osterhoff, B.~Schmidt, V.~Wacker, and A.~R. Maier.
\newblock Chirp mitigation of plasma-accelerated beams by a modulated plasma
  density.
\newblock {\em Phys. Rev. Lett.}, 118:214801, May 2017.

\bibitem{Umstadter1996}
D.~Umstadter, J.~K. Kim, and E.~Dodd.
\newblock Laser injection of ultrashort electron pulses into wakefield plasma
  waves.
\newblock {\em Phys. Rev. Lett.}, 76:2073--2076, March 1996.

\bibitem{McGuffey2010}
C.~McGuffey, A.~G.~R. Thomas, W.~Schumaker, T.~Matsuoka, V.~Chvykov, F.~J.
  Dollar, G.~Kalintchenko, V.~Yanovsky, A.~Maksimchuk, K.~Krushelnick, V.~Yu.
  Bychenkov, I.~V. Glazyrin, and A.~V. Karpeev.
\newblock Ionization induced trapping in a laser wakefield accelerator.
\newblock {\em Phys. Rev. Lett.}, 104:025004, January 2010.

\bibitem{Pak2010}
A.~Pak, K.~A. Marsh, S.~F. Martins, W.~Lu, W.~B. Mori, and C.~Joshi.
\newblock Injection and trapping of tunnel-ionized electrons into
  laser-produced wakes.
\newblock {\em Phys. Rev. Lett.}, 104:025003, January 2010.

\bibitem{STII}
Ming Zeng, Min Chen, Zheng-Ming Sheng, Warren~B. Mori, and Jie Zhang.
\newblock Self-truncated ionization injection and consequent monoenergetic
  electron bunches in laser wakefield acceleration.
\newblock {\em Physics of Plasmas}, 21(3):030701, 2014.

\bibitem{TrojanHorse}
B.~Hidding, G.~Pretzler, J.~B. Rosenzweig, T.~K\"onigstein, D.~Schiller, and
  D.~L. Bruhwiler.
\newblock Ultracold electron bunch generation via plasma photocathode emission
  and acceleration in a beam-driven plasma blowout.
\newblock {\em Phys. Rev. Lett.}, 108:035001, January 2012.

\bibitem{hosking2016fundamental}
R.~J. Hosking.
\newblock {\em Fundamental fluid mechanics and magnetohydrodynamics}.
\newblock Springer, Singapore New York, 2016.

\bibitem{SenGupta1999}
Sudip~Sen Gupta and Predhiman~K. Kaw.
\newblock Phase mixing of nonlinear plasma oscillations in an arbitrary mass
  ratio cold plasma.
\newblock {\em Physical Review Letters}, 82(9):1867--1870, 1999.

\bibitem{Gilljohann2019}
M.{\hspace{0.167em}}F. Gilljohann, H.~Ding, A.~D\"{o}pp, J.~G\"{o}tzfried,
  S.~Schindler, G.~Schilling, S.~Corde, A.~Debus, T.~Heinemann, B.~Hidding,
  S.{\hspace{0.167em}}M. Hooker, A.~Irman, O.~Kononenko, T.~Kurz, A.~Martinez
  de~la Ossa, U.~Schramm, and S.~Karsch.
\newblock Direct observation of plasma waves and dynamics induced by
  laser-accelerated electron beams.
\newblock {\em Physical Review X}, 9(1), 2019.

\bibitem{Semushin2001}
S.~Semushin and V.~Malka.
\newblock High density gas jet nozzle design for laser target production.
\newblock {\em Review of Scientific Instruments}, 72(7):2961--2965, 2001.

\bibitem{COUPERUS2016}
J.P. Couperus, A.~Köhler, T.A.W. Wolterink, A.~Jochmann, O.~Zarini, H.M.J.
  Bastiaens, K.J. Boller, A.~Irman, and U.~Schramm.
\newblock Tomographic characterisation of gas-jet targets for laser wakefield
  acceleration.
\newblock {\em Nuclear Instruments and Methods in Physics Research Section A:
  Accelerators, Spectrometers, Detectors and Associated Equipment}, 830:504 --
  509, 2016.

\bibitem{Kushcel2018}
S.~Kuschel, M.~B. Schwab, M.~Yeung, D.~Hollatz, A.~Seidel, W.~Ziegler,
  A.~S\"avert, M.~C. Kaluza, and M.~Zepf.
\newblock Controlling the self-injection threshold in laser wakefield
  accelerators.
\newblock {\em Phys. Rev. Lett.}, 121:154801, October 2018.

\bibitem{Osterhoff2008}
J.~Osterhoff, A.~Popp, Zs. Major, B.~Marx, T.~P. Rowlands-Rees, M.~Fuchs,
  M.~Geissler, R.~H\"orlein, B.~Hidding, S.~Becker, E.~A. Peralta, U.~Schramm,
  F.~Gr\"uner, D.~Habs, F.~Krausz, S.~M. Hooker, and S.~Karsch.
\newblock Generation of stable, low-divergence electron beams by
  laser-wakefield acceleration in a steady-state-flow gas cell.
\newblock {\em Phys. Rev. Lett.}, 101:085002, August 2008.

\bibitem{Muggli1999}
P.~Muggli, K.~A. Marsh, S.~Wang, C.~E. Clayton, S.~Lee, T.C. Katsouleas, and
  C.~Joshi.
\newblock Photo-ionized lithium source for plasma accelerator applications.
\newblock {\em IEEE Transactions on Plasma Science}, 27(3):791--799, 1999.

\bibitem{Oz2014}
E.~Öz and P.~Muggli.
\newblock A novel rb vapor plasma source for plasma wakefield accelerators.
\newblock {\em Nuclear Instruments and Methods in Physics Research Section A:
  Accelerators, Spectrometers, Detectors and Associated Equipment}, 740:197 --
  202, 2014.
\newblock Proceedings of the first European Advanced Accelerator Concepts
  Workshop 2013.

\bibitem{Dorchies1999}
F.~Dorchies, J.~R. Marqu{\`{e}}s, B.~Cros, G.~Matthieussent, C.~Courtois,
  T.~V{\'{e}}likoroussov, P.~Audebert, J.~P. Geindre, S.~Rebibo, G.~Hamoniaux,
  and F.~Amiranoff.
\newblock Monomode guiding of1016w/cm2laser pulses over 100 rayleigh lengths in
  hollow capillary dielectric tubes.
\newblock {\em Physical Review Letters}, 82(23):4655--4658, 1999.

\bibitem{Kumarappan2005}
V.~Kumarappan, K.~Y. Kim, and H.~M. Milchberg.
\newblock Guiding of intense laser pulses in plasma waveguides produced from
  efficient, femtosecond end-pumped heating of clustered gases.
\newblock {\em Physical Review Letters}, 94(20), May 2005.

\bibitem{Bobrova2001}
N.~A. Bobrova, A.~A. Esaulov, J.-I. Sakai, P.~V. Sasorov, D.~J. Spence,
  A.~Butler, S.~M. Hooker, and S.~V. Bulanov.
\newblock Simulations of a hydrogen-filled capillary discharge waveguide.
\newblock {\em Physical Review E}, 65(1), 2001.

\bibitem{Bobrova2013}
N.~A. Bobrova, P.~V. Sasorov, C.~Benedetti, S.~S. Bulanov, C.~G.~R. Geddes,
  C.~B. Schroeder, E.~Esarey, and W.~P. Leemans.
\newblock Laser-heater assisted plasma channel formation in capillary discharge
  waveguides.
\newblock {\em Physics of Plasmas}, 20(2):020703, 2013.

\bibitem{vanTilborg2017}
J.~van Tilborg, S.{\hspace{0.167em}}K. Barber, H.-E. Tsai,
  K.{\hspace{0.167em}}K. Swanson, S.~Steinke,
  C.{\hspace{0.167em}}G.{\hspace{0.167em}}R. Geddes, A.{\hspace{0.167em}}J.
  Gonsalves, C.{\hspace{0.167em}}B. Schroeder, E.~Esarey,
  S.{\hspace{0.167em}}S. Bulanov, N.{\hspace{0.167em}}A. Bobrova,
  P.{\hspace{0.167em}}V. Sasorov, and W.{\hspace{0.167em}}P. Leemans.
\newblock Nonuniform discharge currents in active plasma lenses.
\newblock {\em Physical Review Accelerators and Beams}, 20(3), 2017.

\bibitem{Lindstrm2018}
C.{\hspace{0.167em}}A. Lindstr{\o}m, E.~Adli, G.~Boyle, R.~Corsini,
  A.{\hspace{0.167em}}E. Dyson, W.~Farabolini, S.{\hspace{0.167em}}M. Hooker,
  M.~Meisel, J.~Osterhoff, J.-H. R\"{o}ckemann, L.~Schaper, and
  K.{\hspace{0.167em}}N. Sjobak.
\newblock Emittance preservation in an aberration-free active plasma lens.
\newblock {\em Physical Review Letters}, 121(19), 2018.

\bibitem{Weingartner2011}
R.~Weingartner, M.~Fuchs, A.~Popp, S.~Raith, S.~Becker, S.~Chou, M.~Heigoldt,
  K.~Khrennikov, J.~Wenz, T.~Seggebrock, B.~Zeitler, Zs. Major, J.~Osterhoff,
  F.~Krausz, S.~Karsch, and F.~Gr\"{u}ner.
\newblock Imaging laser-wakefield-accelerated electrons using miniature
  magnetic quadrupole lenses.
\newblock {\em Physical Review Special Topics - Accelerators and Beams}, 14(5),
  May 2011.

\bibitem{Steinke2016}
S.~Steinke, J.~van Tilborg, C.~Benedetti, C.~G.~R. Geddes, C.~B. Schroeder,
  J.~Daniels, K.~K. Swanson, A.~J. Gonsalves, K.~Nakamura, N.~H. Matlis, B.~H.
  Shaw, E.~Esarey, and W.~P. Leemans.
\newblock Multistage coupling of independent laser-plasma accelerators.
\newblock {\em Nature}, 530(7589):190--193, 2016.

\bibitem{vanTilborg2018}
J.~van Tilborg, S.~K. Barber, C.~Benedetti, C.~B. Schroeder, F.~Isono, H.-E.
  Tsai, C.~G.~R. Geddes, and W.~P. Leemans.
\newblock Comparative study of active plasma lenses in high-quality electron
  accelerator transport lines.
\newblock {\em Physics of Plasmas}, 25(5):056702, May 2018.

\bibitem{atzeni2004}
Stefano Atzeni.
\newblock {\em The physics of inertial fusion : beam plasma interaction,
  hydrodynamics, hot dense matter}.
\newblock Clarendon Press Oxford University Press, Oxford New York, 2004.

\bibitem{gonsalves2011tunable}
AJ~Gonsalves, Kei Nakamura, Chen Lin, Dmitriy Panasenko, Satomi Shiraishi,
  Thomas Sokollik, Carlo Benedetti, CB~Schroeder, CGR Geddes, Jeroen
  Van~Tilborg, et~al.
\newblock Tunable laser plasma accelerator based on longitudinal density
  tailoring.
\newblock {\em Nature Physics}, 7(11):862--866, 2011.

\bibitem{bendoyro2008plasma}
Rodolfo~A Bendoyro, Roxana~I Onofrei, Jo{\~A}o Sampaio, Rita Macedo,
  Gon{\c{C}}alo Figueira, and Nelson~C Lopes.
\newblock Plasma channels for electron accelerators using discharges in
  structured gas cells.
\newblock {\em IEEE transactions on plasma science}, 36(4):1728--1733, 2008.

\bibitem{d2019flashforward}
R~D'Arcy, A~Aschikhin, S~Bohlen, G~Boyle, T~Br{\"u}mmer, J~Chappell,
  S~Diederichs, B~Foster, MJ~Garland, L~Goldberg, et~al.
\newblock Flashforward: plasma wakefield accelerator science for
  high-average-power applications.
\newblock {\em Philosophical Transactions of the Royal Society A},
  377(2151):20180392, 2019.

\bibitem{schaper2014longitudinal}
Lucas Schaper, Lars Goldberg, Tobias Kleinw{\"a}chter, Jan-Patrick
  Schwinkendorf, and Jens Osterhoff.
\newblock Longitudinal gas-density profilometry for plasma-wakefield
  acceleration targets.
\newblock {\em Nuclear Instruments and Methods in Physics Research Section A:
  Accelerators, Spectrometers, Detectors and Associated Equipment},
  740:208--211, 2014.

\bibitem{gigosos1996new}
Marco~A Gigosos and Valentin Cardenoso.
\newblock New plasma diagnosis tables of hydrogen stark broadening including
  ion dynamics.
\newblock {\em Journal of Physics B: Atomic, Molecular and Optical Physics},
  29(20):4795, 1996.

\bibitem{gigosos2003computer}
Marco~A Gigosos, Manuel~A Gonzalez, and Valentin Cardenoso.
\newblock Computer simulated balmer-alpha,-beta and-gamma stark line profiles
  for non-equilibrium plasmas diagnostics.
\newblock {\em Spectrochimica Acta Part B: Atomic Spectroscopy},
  58(8):1489--1504, 2003.

\bibitem{kalal1988abel}
Milan Kalal and Keith Nugent.
\newblock Abel inversion using fast fourier transforms.
\newblock {\em Applied optics}, 27(10):1956--1959, 1988.

\bibitem{van2019density}
Jeroen Van~Tilborg, Anthony Gonsalves, Carl Schroeder, Wim Leemans, Cameron
  Geddes, and Eric Esarey.
\newblock Density characterization of discharged capillaries through
  common-path spectral-domain interferometry.
\newblock {\em Bulletin of the American Physical Society}, 64, 2019.

\bibitem{garland2020characterisation}
JM~Garland, G~Tauscher, L~Schaper, L~Goldberg, K~Poder, and Osterhof J.
\newblock Temporally resolved longitudinal plasma density profile
  characterisation in discharge capillaries for plasma wakefield acceleration.
\newblock {\em To be submitted}, 2020.

\end{thebibliography}
\bibliographystyle{unsrt}

%
%
%
%
%
%
%
\end{document}